\documentclass[11pt]{article}
\DeclareMathAlphabet{\scr}{U}{rsfs}{m}{n}

\pdfoutput=1
\usepackage{latexsym}
\usepackage{epsfig}
\usepackage[mathscr]{eucal}
\usepackage{amsfonts}
\usepackage{amscd}
\usepackage{cite}
\usepackage{array}
\usepackage{amssymb}
\usepackage{colordvi}
\usepackage[centertags]{amsmath}
\usepackage{enumerate}
\usepackage{graphicx}
\usepackage{booktabs}
\usepackage{theorem}
\usepackage[footnotesize]{caption}
\usepackage{soul}
\usepackage{url}
\usepackage[hidelinks]{hyperref}
\usepackage{mcite}
\usepackage{slashed}
\usepackage{color}
\usepackage{bbm}
\usepackage[utf8]{inputenc}
\newcommand{\newc}{\newcommand}
\newc{\be}{\begin{equation}}
\newc{\ee}{\end{equation}}
\newc{\bea}{\begin{eqnarray}}
\newc{\eea}{\end{eqnarray}}
\newc{\ol}{\overline}
\newc{\wt}{\widetilde}
\newc{\bs}{\boldsymbol}
\newc{\m}{\mathcal}
\newc{\la}{\langle}
\newc{\ra}{\rangle}

\newcommand{\beq}{\begin{eqnarray}}
\newcommand{\eeq}{\end{eqnarray}}
\newcommand{\bpmatrix}{\begin{pmatrix}}
\newcommand{\epmatrix}{\end{pmatrix}}
\newcommand{\ba}{\begin{array}}
\newcommand{\ea}{\end{array}}

\renewcommand{\ol}{\text{1l}}

\renewcommand{\eqref}[1]{Eq.~(\ref{#1})}

\newcommand{\bc}{\begin{center}}
\newcommand{\ec}{\end{center}}

\newcommand{\s}{\newline \vspace*{-3.5mm}}

\begin{document}

\title{
\vspace*{-3.7cm}
\phantom{h} \hfill\mbox{\small KA-TP-20-2018}\\[-1.1cm]
\phantom{h} \hfill\mbox{\small DESY 18-130}
\\[1cm]
\textbf{Models with Extended Higgs Sectors at Future $e^+ e^-$ Colliders \\[4mm]}}

\date{}
\author{
Duarte Azevedo$^{1,4\,}$\footnote{E-mail:
\texttt{dazevedo@alunos.fc.ul.pt}} ,
Pedro Ferreira$^{1,2\,}$\footnote{E-mail:
\texttt{pmmferreira@fc.ul.pt}} ,
M.~Margarete M\"uhlleitner$^{3\,}$\footnote{E-mail:
\texttt{milada.muehlleitner@kit.edu}} ,\\
Rui Santos$^{1,2,4\,}$\footnote{E-mail:
  \texttt{rasantos@fc.ul.pt}} ,
Jonas Wittbrodt$^{5\,}$\footnote{E-mail: \texttt{jonas.wittbrodt@desy.de}}
\\[5mm]
{\small\it
$^1$Centro de F\'{\i}sica Te\'{o}rica e Computacional,
    Faculdade de Ci\^{e}ncias,} \\
{\small \it    Universidade de Lisboa, Campo Grande, Edif\'{\i}cio C8
  1749-016 Lisboa, Portugal} \\[3mm]
{\small\it
$^2$ISEL -
 Instituto Superior de Engenharia de Lisboa,} \\
{\small \it   Instituto Polit\'ecnico de Lisboa
 1959-007 Lisboa, Portugal} \\[3mm]
{\small\it
$^3$Institute for Theoretical Physics, Karlsruhe Institute of Technology,} \\
{\small\it 76128 Karlsruhe, Germany}\\[3mm]
{\small\it
$^4$LIP, Departamento de F\'{\i}sica, Universidade do Minho, 4710-057 Braga, Portugal}\\[3mm]
{\small\it
$^5$Deutsches Elektronen-Synchrotron DESY, Notkestra{\ss}e 85, 22607
Hamburg, Germany}
}

\maketitle

\begin{abstract}
We discuss the phenomenology of several Beyond the Standard Model (SM) extensions that include
extended Higgs sectors. The models discussed are: the SM extended by a complex singlet field (CxSM),
the 2-Higgs-Doublet Model with a CP-conserving (2HDM) and a CP-violating (C2HDM) scalar sector,
the singlet extension of the 2-Higgs-Doublet Model (N2HDM), and the Next-to-Minimal Supersymmetric SM extension (NMSSM).
All the above models have at least three neutral scalars, with one being
the 125 GeV Higgs boson. This common feature allows us to compare
the production and decay rates of the other two scalars and therefore to compare their behaviour at future
electron-positron colliders. Using predictions on the
  expected precision of the 125 GeV Higgs
boson couplings at these colliders we are able to obtain
the allowed admixtures of either a singlet or a pseudoscalar
to the observed 125 GeV scalar. Therefore, even if no new scalar is found,
the expected precision at future electron-positron colliders, such as
CLIC, will certainly contribute to a clearer picture of the nature of
the discovered Higgs boson.
\end{abstract}
\thispagestyle{empty}
\vfill
\newpage
\setcounter{page}{1}

\section{Introduction}

The discovery of the Higgs boson by the LHC experiments ATLAS \cite{Aad:2012tfa}
and CMS \cite{Chatrchyan:2012ufa} has triggered the search for new scalars
as predicted by Beyond the Standard Model (BSM) models with extended Higgs sectors.
Although no new scalars were found at the LHC up until now, and no solid hints of new physics
have been reported by the LHC collaborations, the increasing precision in the measurement
of the Higgs couplings to fermions and gauge bosons has dramatically reduced the parameter space
of BSM models. Hence, it could be that at the end of the LHC run we will not discover any new particle
and will have to rely on future colliders to further search for new physics.

In this work we discuss the phenomenology of several BSM extensions that include
extended Higgs sectors at a future electron-positron collider. The models discussed are: the SM extended by a complex singlet field (CxSM),
the 2-Higgs-Doublet Model with a CP-conserving (2HDM) and a CP-violating violating (C2HDM) scalar sector,
the singlet extension of the 2-Higgs-Doublet Model (N2HDM), and the Next-to-Minimal Supersymmetric SM extension (NMSSM).
All the above models have at least three neutral bosons, with one
being the 125 GeV Higgs boson. This common feature allows us to compare
the production and decay rates of the other two scalars.

The models are investigated by performing parameter scans
that take into account the most relevant theoretical and experimental constraints.
Our main goal is to answer two questions. The first one is what can an electron-positron
collider tell us about the nature of the discovered Higgs boson - is it just part of a doublet,
or two doublets; has it a singlet component or a CP-violating one, and if so how large?
The second one is, to what extent can a future electron-positron collider distinguish
between the different BSM versions if a new Higgs boson is found? Are we able to disentangle the models based on Higgs
rate measurements? We hope that we can shed some light on the relevance
of a future electron-positron collider for BSM Higgs searches. This is part (see~\cite{No:2018fev, Buttazzo:2018qqp} for
recents studies on similar subjects) of an effort to build a strong physics case for the next electron-positron colliders.

The outline of the paper is as follows. In section \ref{sec:model} we
briefly introduce the models under study.  In section~\ref{sec:scans} we describe the constraints
on the models and how the scans over the parameter space are performed.
 In section \ref{sec:pheno} we discuss what we can learn about the nature of the
 discovered 125 GeV scalar after CLIC.  In section~\ref{sec:sigrates} the signal rates of the
non-SM-like Higgs bosons are compared within the different models.
Our conclusions are given in
section~\ref{sec:concl}.

\section{Description of the Models \label{sec:model}}
\setcounter{equation}{0}
We start with a very brief description of the models analysed in this work and
we refer the reader to~\cite{Muhlleitner:2017dkd} for a detailed description. Here we will
just set our notation and define the free parameters used in each model.

\subsection{The Complex Singlet Extension of the SM \label{sec:cxsm}}
The first model we discuss is an extension of the SM by a complex scalar field (CxSM)
which is defined by a scalar potential  with a softly broken global $U(1)$ symmetry given by
\beq
V = \frac{m^2}{2} H^\dagger H + \frac{\lambda}{4} (H^\dagger
H)^2+\frac{\delta_2}{2} H^\dagger H |\mathbb{S}|^2 +
\frac{b_2}{2}|\mathbb{S}|^2+ \frac{d_2}{4} |\mathbb{S}|^4 +
\left(\frac{b_1}{4} \mathbb{S}^2 + a_1 \mathbb{S} +c.c. \right)
\, ,  \label{eq:VCxSM}
\eeq
where $\mathbb{S} =  S + iA$ is a hypercharge zero scalar field and the soft breaking
terms are written in parenthesis. We write the fields as
\begin{equation}
H=\left(\begin{array}{c} G^+ \\
   \dfrac{1}{\sqrt{2}} ( v+h+iG^0) \end{array}\right) \quad \mbox{and} \quad
\mathbb{S}=\dfrac{1}{\sqrt{2}}\left[v_S+s+i(v_A+ a)\right] \;,
\label{eq:fieldsCxSM}
\end{equation}
where $v\approx 246$~GeV is the vacuum expectation value (VEV) of the $h$ field and $v_S$ and $v_A$
are the VEVs of the real and imaginary parts of the complex singlet field, respectively.
Except for the soft breaking terms, all parameters are real as required by the hermiticity of the potential. As we further
impose invariance under  $\mathbb{S} \to \mathbb{S}^*$ (or $A \to -A$),  $a_1$ and $b_1$ are real. We choose to work in the broken phase (all three VEVs are non-zero) because
this phase leads to mixing between the three CP-even scalars. Their mass eigenstates  are denoted by $H_i$ and are obtained
from the gauge eigenstates via the rotation matrix $R$ parametrised as
\beq
R =\left( \begin{array}{ccc}
c_{1} c_{2} & s_{1} c_{2} & s_{2}\\
-(c_{1} s_{2} s_{3} + s_{1} c_{3})
& c_{1} c_{3} - s_{1} s_{2} s_{3}
& c_{2} s_{3} \\
- c_{1} s_{2} c_{3} + s_{1} s_{3} &
-(c_{1} s_{3} + s_{1} s_{2} c_{3})
& c_{2}  c_{3}
\end{array} \right) \;,
\label{eq:rotsinglet}
\eeq
where we have defined  $s_{i} \equiv \sin \alpha_i$ and $c_{i} \equiv \cos \alpha_i$, and without loss of generality the angles vary in the range
\beq
-\frac{\pi}{2} \le \alpha_i < \frac{\pi}{2} \,,
\label{eg:alpharanges}
\eeq
and the masses of the neutral Higgs bosons are ordered as $m_{H_1} \leq m_{H_2} \leq m_{H_3}$,
We choose as input parameters the set
\beq
\alpha_1 \;, \quad \alpha_2 \;, \quad \alpha_3 \;, \quad v \;, \quad v_S \;,
\quad m_{H_1} \quad \mbox{and} \quad m_{H_3} \;, \label{eq:cxsminput}
\eeq
and the remaining parameters are determined internally in
{\tt ScannerS} \cite{Coimbra:2013qq,ScannerS} fulfilling the
minimum conditions of the vacuum.

In the broken phase, the couplings
of each Higgs boson, $H_i$, to SM particles are rescaled by a common factor $R_{i1}$.
The expression for all couplings can be found in the
appendix B.1 of \cite{Costa:2015llh}.  All Higgs branching ratios,
including the state-of-the art higher order QCD corrections and
possible off-shell decays can be obtained
from {\tt sHDECAY}\cite{Costa:2015llh}\footnote{The program {\tt sHDECAY} can be downloaded
  from the url: \url{http://www.itp.kit.edu/~maggie/sHDECAY}.}
which is an implementation of
the CxSM and also the RxSM both in their symmetric and broken phases in
{\tt HDECAY}~\cite{Djouadi:1997yw,Djouadi:2018xqq}. A detailed
description of the program can be found in appendix A of \cite{Costa:2015llh}. \s

\subsection{The 2HDM and the C2HDM\label{sec:2hdm}}
In this section we introduce the real (2HDM) and complex (C2HDM) versions of
a particular 2-Higgs-Doublet model, where we add a second doublet to the SM scalar sector.
The Higgs potential is invariant under the $\mathbb{Z}_2$ transformations
$\Phi_1 \to  \Phi_1$ and $\Phi_2 \to - \Phi_2$ and is written as
\beq
V &=& m_{11}^2 |\Phi_1|^2 + m_{22}^2 |\Phi_2|^2 - m_{12}^2 (\Phi_1^\dagger
\Phi_2 + h.c.) + \frac{\lambda_1}{2} (\Phi_1^\dagger \Phi_1)^2 +
\frac{\lambda_2}{2} (\Phi_2^\dagger \Phi_2)^2 \nonumber \\
&& + \lambda_3
(\Phi_1^\dagger \Phi_1) (\Phi_2^\dagger \Phi_2) + \lambda_4
(\Phi_1^\dagger \Phi_2) (\Phi_2^\dagger \Phi_1) +
[\frac{\lambda_5}{2} (\Phi_1^\dagger \Phi_2)^2 + h.c.] \; .
\eeq
By extending the $\mathbb{Z}_2$ symmetry to the fermions we guarantee the absence of
tree-level Flavour Changing Neutral Currents (FCNC). If all parameters
of the potential are real and the VEVs in each doublet are also real
the potential is CP-conserving and we call the model 2HDM; if the VEVs
are real but  $m_{12}^2$ and $\lambda_5$ are complex, with different unrelated phases, the model
is CP-violating and we call it C2HDM~\cite{Ginzburg:2002wt}.
Both the 2HDM and the C2HDM have two charged Higgs bosons
and three neutral scalars. In the 2HDM the neutral scalars are $h$
and $H$, the lighter and the heavier CP-even states, while $A$ is the
CP-odd state. In the C2HDM we have three Higgs
mass eigenstates $H_i$ ($i=1,2,3$) with no definite CP and that are ordered
by ascending mass according to $m_{H_1} \le m_{H_2} \le m_{H_3}$. The rotation
matrix,  $R$, that diagonalises the mass matrix is parametrised as defined for the complex
singlet extension case in Eq.~(\ref{eq:rotsinglet}) and with the same range as in Eq.~(\ref{eg:alpharanges})
for the mixing angles. The CP-conserving 2HDM is obtained from the C2HDM by setting $\alpha_2 = \alpha_3= 0$ and
$\alpha_1= \alpha + \pi/2$ \cite{Khater:2003wq}. In this case the CP-even mass
eigenstates $h$ and $H$ are obtained from the gauge eigenstates through the
rotation parametrised in terms of the angle $\alpha$.
The 2HDM has eight independent parameters while the C2HDM has nine independent parameters. We define for both versions of the model
$v = \sqrt{v_1^2 + v_2^2}  \approx 246\mbox{ GeV}$ and $\tan \beta=v_2/v_1$. For the 2HDM we
choose as independent parameters
\beq
v  \;, \quad \tan\beta \;, \quad \alpha \;, \quad m_{h} \;, \quad m_{H} \;, \quad m_{A} \;, \quad m_{H^\pm} \quad
\mbox{and} \quad m_{12}^2 \;,
\label{eq:2hdminputset}
\eeq
while for the C2HDM we choose~\cite{ElKaffas:2007rq}
\beq
v  \;, \quad \tan\beta \;, \quad \alpha_{1,2,3}
\;, \quad m_{H_i} \;, \quad m_{H_j} \;, \quad m_{H^\pm} \quad
\mbox{and} \quad \mbox{Re}(m_{12}^2) \;,
\label{eq:c2hdminputset}
\eeq
where $m_{H_i}$ and $m_{H_j}$ denote any two of the three neutral Higgs
bosons but where one of them is the 125 GeV scalar. The remaining mass is obtained from the other parameters \cite{ElKaffas:2007rq}.

We write the couplings to massive gauge bosons ($V=W,Z$) of the Higgs boson $H_i$ in the C2HDM as
\beq
i \, g_{\mu\nu} \, c(H_i VV) \, g_{H^{\text SM} VV} \;, \label{eq:gaugecoupdef}
\eeq
where~\cite{Fontes:2014xva}
\beq
c(H_i VV) = c_\beta R_{i1} + s_\beta R_{i2} \;, \label{eq:c2dhmgaugecoup}
\eeq
and $g_{H^{\text SM} VV}$ denotes the SM Higgs coupling factors. In terms of
the gauge boson masses $M_W$ and $M_Z$, the $SU(2)_L$ gauge coupling
$g$ and the Weinberg angle $\theta_W$ they are given by $g_{H^{\text SM} VV} = g M_W$ for $V=W$ and $g M_Z /\cos\theta_W$ for $V=Z$.

Both the 2HDM and C2HDM are free from tree-level FCNCs
by extending the global $\mathbb{Z}_2$ symmetry to the Yukawa sector. The four independent $\mathbb{Z}_2$ charge assignments of the fermion fields
determine the four types of 2HDMs depicted
in Table~\ref{tab:types}.
\begin{table}
\begin{center}
\begin{tabular}{rccc} \toprule
& $u$-type & $d$-type & leptons \\ \midrule
Type I & $\Phi_2$ & $\Phi_2$ & $\Phi_2$ \\
Type II & $\Phi_2$ & $\Phi_1$ & $\Phi_1$ \\
Lepton-specific & $\Phi_2$ & $\Phi_2$ & $\Phi_1$ \\
Flipped & $\Phi_2$ & $\Phi_1$ & $\Phi_2$ \\ \bottomrule
\end{tabular}
\caption{The four Yukawa types of the $\mathbb{Z}_2$-symmetric 2HDM
  defined by the Higgs doublet that couples to each kind of fermions. \label{tab:types}}
\end{center}
\end{table}
The Yukawa Lagrangian is defined by
\beq
{\cal L}_Y = - \sum_{i=1}^3 \frac{m_f}{v} \bar{\psi}_f \left[ c^e(H_i
  ff) + i c^o(H_i ff) \gamma_5 \right] \psi_f H_i \;, \label{eq:yuklag}
\eeq
where $\psi_f$ is the fermion field with mass $m_f$. In Table~\ref{tab:yukcoup} we present the CP-even and the CP-odd components of the Yukawa
couplings, $c^e(H_i ff)$ and $c^o (H_i ff)$, respectively~\cite{Fontes:2014xva}.
\begin{table}
\begin{center}
\begin{tabular}{rccc} \toprule
& $u$-type & $d$-type & leptons \\ \midrule
Type I & $\frac{R_{i2}}{s_\beta} - i \frac{R_{i3}}{t_\beta} \gamma_5$
& $\frac{R_{i2}}{s_\beta} + i \frac{R_{i3}}{t_\beta} \gamma_5$ &
$\frac{R_{i2}}{s_\beta} + i \frac{R_{i3}}{t_\beta} \gamma_5$ \\
Type II & $\frac{R_{i2}}{s_\beta} - i \frac{R_{i3}}{t_\beta} \gamma_5$
& $\frac{R_{i1}}{c_\beta} - i t_\beta R_{i3} \gamma_5$ &
$\frac{R_{i1}}{c_\beta} - i t_\beta R_{i3} \gamma_5$ \\
Lepton-specific & $\frac{R_{i2}}{s_\beta} - i \frac{R_{i3}}{t_\beta} \gamma_5$
& $\frac{R_{i2}}{s_\beta} + i \frac{R_{i3}}{t_\beta} \gamma_5$ &
$\frac{R_{i1}}{c_\beta} - i t_\beta R_{i3} \gamma_5$ \\
Flipped & $\frac{R_{i2}}{s_\beta} - i \frac{R_{i3}}{t_\beta} \gamma_5$
& $\frac{R_{i1}}{c_\beta} - i t_\beta R_{i3} \gamma_5$ &
$\frac{R_{i2}}{s_\beta} + i \frac{R_{i3}}{t_\beta} \gamma_5$ \\ \bottomrule
\end{tabular}
\caption{Components of the Yukawa couplings of the Higgs
  bosons $H_i$ in the C2HDM. The expressions correspond to
  $[c^e(H_i ff) +i c^o (H_i ff) \gamma_5]$ from
  Eq.~(\ref{eq:yuklag}) and $t_\beta$ stands for $\tan \beta$. \label{tab:yukcoup}}
\end{center}
\end{table}
All Higgs branching ratios can be obtained
from {\tt C2HDM\_HDECAY}\cite{Fontes:2017zfn}\footnote{The program {\tt C2HDM\_HDECAY} can be downloaded
from the url: \url{https://www.itp.kit.edu/~maggie/C2HDM}.}
which implements
the C2HDM in {\tt HDECAY}~\cite{Djouadi:1997yw,Djouadi:2018xqq}. These
include state-of-the art higher order QCD corrections and possible off-shell decays.
The complete set of Feynman rules for the C2HDM  is available at: \\[0.1cm]
\centerline{\tt http://porthos.tecnico.ulisboa.pt/arXiv/C2HDM/}

\vspace*{0.1cm} \noindent
where for the SM subset the notation for the covariant derivatives
is the one in~\cite{Romao:2012pq} with all $\eta$'s positive, where the $\eta$'s define the sign of the covariant derivative (see~\cite{Romao:2012pq}).
Note that the 2HDM branching ratios
are part of the  {\tt HDECAY} release (see \cite{Djouadi:1997yw,Djouadi:2018xqq,Harlander:2013qxa} for details).

\subsection{The N2HDM\label{sec:n2hdm}}
The version of the N2HDM used in this work was discussed in great detail in~\cite{Muhlleitner:2016mzt}.
This extension consists of the addition of an extra doublet and an extra real singlet to the SM field content.
The potential is invariant under two
discrete  $\mathbb{Z}_2$ symmetries. The first $\mathbb{Z}_2$  symmetry is just a generalisation of the one used for the 2HDM
in order to avoid tree-level FCNCs,
\begin{align}
  \Phi_1 \to \Phi_1\;, \quad \Phi_2 \to - \Phi_2\;, \quad \Phi_S \to \Phi_S \label{eq:2HDMZ2}
\end{align}
and that is softly broken by $m_{12}^2$;  the second one is defined as
\begin{align}
  \Phi_1 \to \Phi_1\;, \quad \Phi_2 \to \Phi_2\;, \quad \Phi_S \to -\Phi_S \label{eq:singZ2}
\end{align}
and it is not explicitly broken.  $\Phi_1$ and $\Phi_2$ are doublet fields and $\Phi_S$ is a singlet field. The most general form of this
scalar potential invariant under the above transformations is\footnote{Another version of the N2HDM with a different discrete symmetry
was considered in~\cite{Azevedo:2018fmj}. That model allows a dark matter candidate and CP-violation in the dark sector.}
\beq
V &=& m_{11}^2 |\Phi_1|^2 + m_{22}^2 |\Phi_2|^2 - m_{12}^2 (\Phi_1^\dagger
\Phi_2 + h.c.) + \frac{\lambda_1}{2} (\Phi_1^\dagger \Phi_1)^2 +
\frac{\lambda_2}{2} (\Phi_2^\dagger \Phi_2)^2 \nonumber \\
&& + \lambda_3
(\Phi_1^\dagger \Phi_1) (\Phi_2^\dagger \Phi_2) + \lambda_4
(\Phi_1^\dagger \Phi_2) (\Phi_2^\dagger \Phi_1) + \frac{\lambda_5}{2}
[(\Phi_1^\dagger \Phi_2)^2 + h.c.] \nonumber \\
&& + \frac{1}{2} m_S^2 \Phi_S^2 + \frac{\lambda_6}{8} \Phi_S^4 +
\frac{\lambda_7}{2} (\Phi_1^\dagger \Phi_1) \Phi_S^2 +
\frac{\lambda_8}{2} (\Phi_2^\dagger \Phi_2) \Phi_S^2 \;.
\label{eq:n2hdmpot}
\eeq
The doublet and singlet fields after electroweak symmetry breaking can be parametrised as
\beq
\Phi_1 = \left( \begin{array}{c} \phi_1^+ \\ \frac{1}{\sqrt{2}} (v_1 +
    \rho_1 + i \eta_1) \end{array} \right) \;, \quad
\Phi_2 = \left( \begin{array}{c} \phi_2^+ \\ \frac{1}{\sqrt{2}} (v_2 +
    \rho_2 + i \eta_2) \end{array} \right) \;, \quad
\Phi_S = v_S + \rho_S \;,
\eeq
where $v_{1,2}$ are the VEVs of the doublets $\Phi_1$ and $\Phi_2$, respectively, and $v_S$ is
the singlet VEV. The singlet VEV breaks the second $\mathbb{Z}_2$
symmetry, precluding the existence of a dark matter candidate.
As this is a CP-conserving model, with no dark matter candidate, we end up
with three CP-even scalars, one of which plays the role of the 125 GeV
Higgs boson,
a CP-odd scalar and two charged scalars. The orthogonal matrix $R$ that diagonalises
the mass matrix is again parametrised as in Eq.~(\ref{eq:rotsinglet}) in terms of the mixing angles
$\alpha_i$ with the same ranges as before, see \eqref{eg:alpharanges}.
The physical CP-even eigenstates, denoted by $H_1$, $H_2$ and $H_3$, are ordered by ascending mass as
\beq
m_{H_1} < m_{H_2} < m_{H_3} \;.
\eeq
We choose as the 12 independent parameters the set
\beq
\alpha_1 \; , \quad \alpha_2 \; , \quad \alpha_3 \; , \quad t_\beta \;
, \quad v \; ,
\quad v_s \; , \quad m_{H_{1,2,3}} \;, \quad m_A \;, \quad m_{H^\pm}
\;, \quad m_{12}^2 \;.
\eeq
The expressions of the quartic couplings in terms of
the physical parameter set can be found in appendix~A.1 of
\cite{Muhlleitner:2016mzt}. All Higgs branching ratios, including the
state-of-the art higher order QCD corrections and possible off-shell
decays can be obtained
from {\tt N2HDECAY}\footnote{The program {\tt N2HDECAY} is available at: \url{https://gitlab.com/jonaswittbrodt/N2HDECAY.}
}~\cite{Muhlleitner:2016mzt,Engeln:2018mbg} which implements the N2HDM
 in {\tt HDECAY}~\cite{Djouadi:1997yw,Djouadi:2018xqq}.\s

\subsection{The NMSSM \label{sec:nmssm}}
Supersymmetric models require the introduction of at least two Higgs
doublets. The NMSSM extends the two Higgs doublet superfields $\hat{H}_u$ and
$\hat{H}_d$ of the Minimal Supersymmetric
extension (MSSM) by a complex superfield $\hat{S}$. The $\mu$ problem
of the MSSM is thus solved dynamically
when the singlet field acquires a non-vanishing VEV. The NMSSM Higgs
sector consists of seven physical Higgs states after EWSB. These are,
in the CP-conserving case investigated in this work, three neutral
CP-even, two neutral CP-odd ones and a pair of charged Higgs
bosons. The NMSSM Higgs potential is derived from the superpotential,
the soft SUSY breaking Lagrangian and the $D$-term contributions. The
scale-invariant NMSSM superpotential reads in terms of the hatted superfields
\beq
{\cal W} = \lambda \widehat{S} \widehat{H}_u \widehat{H}_d +
\frac{\kappa}{3} \, \widehat{S}^3 + h_t
\widehat{Q}_3\widehat{H}_u\widehat{t}_R^c - h_b \widehat{Q}_3
\widehat{H}_d\widehat{b}_R^c  - h_\tau \widehat{L}_3 \widehat{H}_d
\widehat{\tau}_R^c \; .
\label{eq:superpotential}
\eeq
For simplicity, we have only included the third generation fermion
superfields here. They are given by the left-handed doublet quark
($\widehat{Q}_3$) and lepton ($\widehat{L}_3$) superfields and the
right-handed singlet quark ($\widehat{t}^c_R,\widehat{b}^c_R$) and
lepton ($\widehat{\tau}^c_R$) superfields. The first term in Eq.~(\ref{eq:superpotential})
takes the role of the $\mu$-term $\mu \hat{H}_d \hat{H}_u$
of the MSSM superpotential, the term cubic in the singlet superfield
breaks the Peccei-Quinn symmetry thus avoiding a
massless axion and the last three terms represent the Yukawa
interactions. The soft SUSY breaking Lagrangian consists of
the mass terms for the Higgs and the sfermion
fields, that are built from the complex scalar components of the superfields,
\beq
\label{eq:Lagmass}
 -{\cal L}_{\mathrm{mass}} &=&
 m_{H_u}^2 | H_u |^2 + m_{H_d}^2 | H_d|^2 + m_{S}^2| S |^2 \nonumber \\
  &+& m_{{\tilde Q}_3}^2|{\tilde Q}_3^2| + m_{\tilde t_R}^2 |{\tilde t}_R^2|
 +  m_{\tilde b_R}^2|{\tilde b}_R^2| +m_{{\tilde L}_3}^2|{\tilde L}_3^2| +
 m_{\tilde  \tau_R}^2|{\tilde \tau}_R^2|\; .
\eeq
The contribution to the
soft SUSY breaking part from the trilinear soft SUSY breaking
interactions between the sfermions and the Higgs fields reads
\beq
\label{eq:Trilmass}
-{\cal L}_{\mathrm{tril}}=  \lambda A_\lambda H_u H_d S + \frac{1}{3}
\kappa  A_\kappa S^3 + h_t A_t \tilde Q_3 H_u \tilde t_R^c - h_b A_b
\tilde Q_3 H_d \tilde b_R^c - h_\tau A_\tau \tilde L_3 H_d \tilde \tau_R^c
+ \mathrm{h.c.} \;
\eeq
where the $A$'s denote the soft SUSY breaking trilinear couplings.
The gaugino mass parameters $M_{1,2,3}$ of
the bino ($\tilde{B}$), winos ($\tilde{W}$) and gluinos ($\tilde{G}$),
respectively, that contribute to the soft SUSY breaking are summarised in
\beq
-{\cal L}_\mathrm{gauginos}= \frac{1}{2} \bigg[ M_1 \tilde{B}
\tilde{B}+M_2 \sum_{a=1}^3 \tilde{W}^a \tilde{W}_a +
M_3 \sum_{a=1}^8 \tilde{G}^a \tilde{G}_a  \ + \ {\rm h.c.}
\bigg] \;.
\eeq
We will allow for non-universal soft terms at the GUT scale.

The expansion of the tree-level scalar potential around the
non-vanishing VEVs of the Higgs doublet and singlet fields,
\beq
H_d = \left( \begin{array}{c} (v_d + h_d + i a_d)/\sqrt{2} \\
   h_d^- \end{array} \right) \,, \;
H_u = \left( \begin{array}{c} h_u^+ \\ (v_u + h_u + i a_u)/\sqrt{2}
 \end{array} \right) \,, \;
S= \frac{v_s+h_s+ia_s}{\sqrt{2}}
\eeq
leads to the Higgs mass matrices for the three scalars ($h_d, h_u,
h_s$), the three pseudoscalars ($a_d,a_u,a_s$) and the charged Higgs
states ($h_u^\pm,h_d^\mp$) that are obtained from the second derivative of
the scalar potential. The VEVs $v_u, v_d$ and $v_s$ are chosen to be
real and positive. Rotation with the orthogonal matrix ${\cal R}^S$
that diagonalises the $3\times 3$ mass matrix squared, $M^2_S$, of
the CP-even fields,
yields the CP-even mass eigenstates $H_i$ ($i=1,2,3$),
\beq
(H_1, H_2, H_3)^T = {\cal R}^S (h_d,h_u,h_s)^T \;.
\label{eq:scalarrotation}
\eeq
They are ordered by ascending mass,
$M_{H_1} \le M_{H_2} \le M_{H_3}$. The CP-odd mass eigenstates $A_1$
and $A_2$ result from a rotation ${\cal R}^G$
separating the massless Goldstone boson followed by a rotation ${\cal R}^P$
into the mass eigenstates,
\beq
(A_1,A_2,G)^T = {\cal R}^P {\cal R}^G (a_d,a_u,a_s)^T \;,
\label{eq:pseudorot}
\eeq
which are ordered by ascending mass, $M_{A_1} \le M_{A_2}$, too.

The three minimisation conditions of the scalar potential are used to
replace the soft SUSY breaking masses squared for $H_u$, $H_d$ and $S$
in ${\cal L}_{\text{mass}}$ by the remaining parameters of the
tree-level scalar potential. This leads to the following six
parameters parametrising the tree-level NMSSM Higgs sector,
\beq
\lambda\ , \ \kappa\ , \ A_{\lambda} \ , \ A_{\kappa}, \
\tan \beta =v_u/ v_d \quad \mathrm{and}
\quad \mu_\mathrm{eff} = \lambda v_s/\sqrt{2}\; .
\eeq
We have chosen the sign conventions such that $\lambda$ and $\tan\beta$
are positive, whereas $\kappa, A_\lambda, A_\kappa$ and
$\mu_{\text{eff}}$ are allowed to have both signs. Contrary to the
non-SUSY Higgs sector extensions introduced in the previous sections,
the Higgs boson masses are not input parameters. They are instead
calculated from these, including higher order corrections. These are
crucial to shift the mass of the SM-like Higgs boson to the observed
value of 125~GeV. Due to these corrections also the
soft SUSY breaking mass terms for the scalars and the gauginos as well
as the trilinear soft SUSY breaking couplings contribute to the Higgs
sector.

\section{Parameter Scans \label{sec:scans}}

The analyses are performed with points, each corresponding to a set of the parameters
chosen for a given model, that are in agreement with the theoretical and experimental constraints.
The discovered SM-like Higgs boson mass is taken to be~\cite{Aad:2015zhl}
\beq
m_{h_{125}} = 125.09 \; \mbox{GeV}\;,
\eeq
and we suppress interfering Higgs signals by forcing any other neutral scalar to be outside the
$m_{h_{125}} \pm 5$~GeV mass window. Any of the Higgs bosons is allowed to be the discovered
one except for charged and pure pseudoscalar particles.
The vacuum expectation value $v$ is fixed by the
$W$ boson mass and all calculations of cross sections and branching ratios do not include electroweak
corrections as they are not fully available for all models.
All models except for the NMSSM, the scan of which will be described
below, have been implemented
as {\tt ScannerS} model classes. This  allowed us to perform a full parameter scan that simultaneously
applies the constraints we will now briefly describe. The theoretical bounds are common to
all models although with different expressions. We force all potentials to be bounded from
below, we require that perturbative unitarity holds and that the electroweak vacuum is the global minimum (using the discriminant from\cite{Ivanov:2015nea} for the C2HDM).

Compatibility with electroweak precision data for the CxSM was imposed by a 95\% C.L.~exclusion
limit from the electroweak precision observables $S$, $T$ and $U$ ~\cite{Peskin:1991sw,Maksymyk:1993zm} -- see
\cite{Costa:2014qga} for more details. The same constraints for the C2HDM use the expressions
in~\cite{Branco:2011iw} while for the N2HDM we use the formulae in~\cite{Grimus:2007if,Grimus:2008nb}.
For the computed values of $S$, $T$ and $U$ we ask for a $2\sigma$ compatibility with the SM fit \cite{Baak:2014ora}
taking into account the full correlation among the three parameters.

95\% C.L.~exclusion limits on non-observed scalars have been applied by using {\tt HiggsBounds} \cite{Bechtle:2013wla}
which include LEP, Tevatron and up-to-date LHC experimental data. Compatibility with the Higgs data
is enforced using the individual signal strengths fit~\cite{Khachatryan:2016vau} for the $h_{125}$.
The branching ratios for the different models were calculated using the modified versions of {\tt HDECAY}
as described in the previous sections. All scalar production cross sections can be easily obtained
from the corresponding SM one except for the gluon fusion ($ggF$) and $b$-quark
fusion ($bbF$) which were determined using  {\tt SusHi  v1.6.0}
\cite{Harlander:2012pb,Harlander:2016hcx}.
For the C2HDM, the CP-even and the CP-odd Yukawa coupling contributions are calculated separately and then added
incoherently, giving
\beq
\mu_F = \frac{\sigma^{\text{even}}_{\text{C2HDM}} (ggF)
  +\sigma^{\text{even}}_{\text{C2HDM}} (bbF)
  +\sigma^{\text{odd}}_{\text{C2HDM}} (ggF) +
  \sigma^{\text{odd}}_{\text{C2HDM}}
  (bbF)}{\sigma^{\text{even}}_{\text{SM}} (ggF)} \;,
\eeq
where we neglected the $bbF$ contribution for the SM in the denominator.
Analogous expressions were used for the other models which do not have a CP-odd component.

 Models with two doublets with or without extra neutral singlets always have a pair of charged Higgs bosons. In this study
 the charged Higgs Yukawa couplings are always proportional to two parameters only: the charged Higgs mass and
 $\tan \beta$. These couplings are constrained by the measurements of $R_b$~\cite{Haber:1999zh,Deschamps:2009rh} and
$B \to X_s \gamma$ \cite{Deschamps:2009rh,Mahmoudi:2009zx,Hermann:2012fc,Misiak:2015xwa,Misiak:2017bgg},
which yields  $2\sigma$ exclusion bounds on the $m_{H^\pm}-t_\beta$ plane. The latest calculation
of~\cite{Misiak:2017bgg} enforces, almost independently of the value of $\tan \beta$,
\beq
m_{H^\pm} > 580 \mbox{ GeV}
\eeq
in the Type II and Flipped models while in Type I and Lepton Specific models  this
bound is not only much weaker but it has a much stronger dependence on $\tan\beta$.

Finally there are bounds that apply only to the C2HDM because constraints on CP violation
in the Higgs sector arise from electric dipole moment (EDM) measurements. Among these the
EDM of the electron imposes the strongest constraints~\cite{Inoue:2014nva}, with the
experimental limit given by the ACME collaboration~\cite{Baron:2013eja}. We require our results
to be compatible with the values given in~\cite{Baron:2013eja} at 90\% C.L. A detailed discussion
of the constraints specific to the C2HDM can be found in~\cite{Fontes:2017zfn}.
With all the above constraints taken into account, the initial range of parameters
chosen for each model is as follows:
\begin{itemize}

\item
\textbf{The CxSM Parameter Range Scan}

The non-125 GeV Higgs bosons are chosen to be in the range
\beq
30 \; \mbox{GeV } \le m_{H_i} < 1 \;\mbox{TeV} ,\;H_i \ne h_{125} \;.
\eeq
The VEVs $v_A$ and $v_S$ are varied in the range
\beq
1 \mbox{ GeV } \le v_A, v_S < 1.5 \mbox{ TeV} \;.
\eeq
The mixing angles $\alpha_{1,2,3}$ vary within the limits
\beq
-\frac{\pi}{2} \le \alpha_{1,2,3} < \frac{\pi}{2} \;.
\eeq

\item
\textbf{The (C)2HDM Parameter Range Scan}

The angles vary in the range
\beq
0.5 \le t_\beta \le 35 \label{eq:tbscanc2hdm}
\eeq
and
\beq
- \frac{\pi}{2} \le \alpha_{1,2,3} < \frac{\pi}{2} \;.
\eeq
The value of $\mbox{Re} (m_{12}^2)$ is in the range
\beq
0 \mbox{ GeV}^2 \le \mbox{Re}(m_{12}^2)  < 500 000 \mbox{ GeV}^2 \;.
\eeq
In
type II, the charged Higgs mass is chosen in the range
\beq
580 \mbox{ GeV } \le m_{H^\pm} < 1 \mbox{ TeV } \;,
\eeq
while in type I
\beq
80 \mbox{ GeV } \le m_{H^\pm} < 1 \mbox{ TeV } \;.
\eeq
The electroweak precision constraints combined with perturbative unitarity bounds force the mass of at
least one of the neutral Higgs bosons to be close to $m_{H^\pm}$.
In order to increase the efficiency of the parameter scan, due to electroweak precision constraints,
the second neutral Higgs mass $m_{H_i \ne h_{125}}$ is in the interval
\beq
500 \mbox{ GeV}\leq m_{H_i}<1 \mbox{ TeV}
\eeq
in type II and
\beq
30 \mbox{ GeV}\leq m_{H_i}<1 \mbox{ TeV}
\eeq
in type I.
In our parametrisation the third neutral Higgs boson $m_{H_j \ne H_i, h_{125}}$
is calculated by {\tt ScannerS} since it is not an independent parameter.

\item
\textbf{The N2HDM Parameter Range Scan}

In view of what was discussed for the previous models, the ranges for the parameters of the N2HDM are
\beq
\begin{array}{ll}
-\frac{\pi}{2} \le \alpha_{1,2,3} < \frac{\pi}{2}\;, &
0.25 \le t_\beta \le 35\;, \\[0.2cm]
0 \mbox{ GeV}^2 \le \mbox{Re}(m_{12}^2) < 500000 \mbox{ GeV}^2\;, &
1 \mbox{ GeV} \le v_S \le 1.5 \mbox{ TeV}\;, \\[0.2cm]
30 \mbox{ GeV} \le m_{H_i \ne m_{h_{125}}}, m_A \le 1 \mbox{ TeV}\;, \\[0.2cm]
80 \mbox{ GeV} \le m_{H^\pm} < 1 \mbox{ TeV (type I)}\;, &
580 \mbox{ GeV} \le m_{H^\pm} < 1 \mbox{ TeV (type II)} \;.
\end{array}
\eeq
\end{itemize}
Note that the 125 GeV Higgs boson can be the lighter as well as the heavier scalar. This possibility
is not excluded in any of the models.

\subsection{The NMSSM Parameter Scan  \label{sec:nmssmscan}}
For the NMSSM parameter scan we proceed as described in
\cite{Costa:2015llh,King:2014xwa} and shortly summarise the main
features. We use the {\tt NMSSMTools} package
\cite{Ellwanger:2004xm,Ellwanger:2005dv,Ellwanger:2006rn,Das:2011dg,Muhlleitner:2003vg,Belanger:2005kh}
to calculate the spectrum of the Higgs and SUSY particles with higher order
corrections included. The package also checks for
the constraints from low-energy observables. It provides the input required by {\tt
  HiggsBounds} which verifies compatibility with the exclusion bounds from
Higgs searches. The relic
density obtained through an interface with {\tt micrOMEGAS}
\cite{Belanger:2005kh} is required not to exceed the value
measured by the PLANCK collaboration \cite{Ade:2013zuv}.
The spin-independent nucleon-dark matter direct
detection cross section, that is also obtained from {\tt micrOMEGAS}, is
required not to violate the upper bound from the LUX
experiment~\cite{Akerib:2016vxi}.
We furthermore test for compatibility with the direct detection limits from XENON1T
\cite{Aprile:2018dbl}.
The mass of one of the neutral
CP-even Higgs bosons has to lie  between 124 and 126 GeV. The
signal strengths of this Higgs boson have to be in agreement with the
signal strength fit of \cite{Khachatryan:2016vau} at the
$2 \times 1\sigma$ level. For the production cross sections,
 gluon fusion and $b\bar{b}$ annihilation, we take the SM cross
sections and multiply them with the effective couplings
obtained from {\tt NMSSMTools}. The SM cross section values are obtained from {\tt
  SusHi} \cite{Harlander:2012pb,Harlander:2016hcx}. In gluon fusion the
next-to-leading order (NLO) corrections are included with the full top quark mass
dependence \cite{Spira:1995rr} and the next-to-next-to-leading order (NNLO)
corrections in the heavy quark effective theory \cite{Harlander:2002wh,Anastasiou:2002yz,Harlander:2002vv,Anastasiou:2002wq,Ravindran:2003um}. For Higgs
masses below 300~GeV the
next-to-next-to-next-to-leading order (N$^3$LO) corrections are taken
into account in a threshold expansion
\cite{Anastasiou:2014lda,Anastasiou:2015yha,Anastasiou:2016cez,Mistlberger:2018etf}. For
masses above 50~GeV $b\bar{b}$ annihilation cross sections that match
between the five- and four-flavor scheme are used obtained in the soft-collinear
effective theory \cite{Bonvini:2015pxa,Bonvini:2016fgf}. They equal
the results from \cite{Forte:2015hba,Forte:2016sja}.  For masses below
50~GeV, cross sections obtained in the Santander matching
\cite{Harlander:2011aa} are used, with the five-flavor scheme cross
sections from \cite{Harlander:2003ai} and the four-flavor scheme ones
from \cite{Dittmaier:2003ej,Dawson:2003kb,Wiesemann:2014ioa}.
The branching ratios are obtained from {\tt NMSSMTools}.
We cross-checked the Higgs branching ratios of {\tt
  NMSSMTools} against {\tt NMSSMCALC} \cite{Baglio:2013iia}.
We demand the masses of all Higgs bosons to be separated by at least
1~GeV in order to avoid overlapping signals.
The obtained parameter points are also checked for
compatibility with the SUSY searches at LHC. We require the gluino
mass and the lightest squark mass of the second
generation to be above 1.85~TeV, respectively,
\cite{Aad:2015iea}. The stops have to be heavier than 800~GeV
\cite{Aaboud:2016lwz} and the slepton masses heavier than 400~GeV
\cite{Aad:2014vma}. The absolute value of the chargino mass must not
be lighter than 300~GeV \cite{Aad:2015jqa}.

The scan ranges applied for the various parameters are summarised in
Table~\ref{tab:nmssmscan}. Perturbativity is ensured by applying
the rough constraint
\beq
\lambda^2 + \kappa^2 < 0.7^2 \;.
\eeq
The remaining mass parameters of the third generation sfermions that
are not listed in the table are chosen as
\beq
m_{\tilde{t}_R} = m_{\tilde{Q}_3} \;, \quad m_{\tilde{\tau}_R} =
m_{\tilde{L}_3} \quad \mbox{ and } \; m_{\tilde{b}_R} = 3 \mbox{ TeV} \;.
\eeq
The mass parameters of the first and second generation sfermions are
set to 3 TeV. For consistency with the parameter ranges of the other
models we kept only points with all Higgs masses between 30~GeV and
1~TeV.
\begin{table}
\begin{center}
\begin{tabular}{l|ccc|cccccccccccc} \toprule
& $t_\beta$ & $\lambda$ & $\kappa$ & $M_1$ & $M_2$ & $M_3$ & $A_t$ &
$A_b$ & $A_\tau$ & $m_{\tilde{Q}_3}$ & $m_{\tilde{L}_3}$ & $A_\lambda$
& $A_\kappa$ & $\mu_{\text{eff}}$ \\
& & & & \multicolumn{11}{c}{in TeV} \\ \midrule
min & 1 & 0 & -0.7 & 0.1 & 0.2 & 1.3 & -6 & -6 & -3 & 0.6 & 0.6 & -2 &
-2 & -5 \\
max & 50 & 0.7 & 0.7 & 1 & 2 & 7 & 6 & 6 & 3 & 4 & 4 & 2 & 2 & 5 \\ \bottomrule
\end{tabular}
\caption{Input parameters for the NMSSM scan. All parameters have been
varied independently between the given minimum and maximum values. \label{tab:nmssmscan}}
\end{center}
\end{table}

\section{Phenomenological Analysis \label{sec:pheno}}

\subsection{The Nature of the 125 GeV Higgs Boson after CLIC}

\begin{table}[h!]\centering
\begin{tabular}{lrrr}
\toprule
Parameter & \multicolumn{3}{c}{Relative precision \cite{Sicking:2016zcl,Abramowicz:2013tzc}}\\
\midrule
                                 & 350~ GeV   & $+1.4$ TeV    & $+3.0$ TeV      \\
                                 & 500 fb$^{-1}$ &  +1.5 ab$^{-1}$  &  +2.0  ab$^{-1}$ \\
\midrule
$\kappa_{HZZ}$             &  0.43\%       &  0.31\%         & 0.23\%        \\
$\kappa_{HWW}$             &  1.5\%        &  0.15\%         & 0.11\%        \\
$\kappa_{Hbb}$           &  1.7\%        &  0.33\%         & 0.21\%        \\
$\kappa_{Hcc}$           &  3.1\%        &  1.1\%          & 0.75\%         \\
$\kappa_{Htt}$           &  $-$         &  4.0\%          & 4.0\%         \\
$\kappa_{H \tau \tau}$           &  3.4\%          & 1.3\%          & $<$1.3\%         \\
$\kappa_{H \mu \mu}$           &  $-$          &  14\%          & 5.5\%  \\
$\kappa_{H gg}$             &  3.6\%         &  0.76\%          & 0.54\%         \\
$\kappa_{H \gamma \gamma}$           &  $-$          &  5.6\%          & $<$ 5.6\%       \\
\bottomrule
\end{tabular}
\caption{Results of the model-dependent global Higgs fit
on the expected precisions of the $\kappa_{Hii}$ (see text). Entries marked ``$-$'' cannot be measured with sufficient precision at the given energy.
We call the first (350 GeV) scenario $Sc1$, the second (1.4 TeV) $Sc2$ and the third (3.0 TeV) $Sc3$.}
 \label{tab:higgs:params}
\end{table}

Over the last years, predictions for the measurement of the Higgs couplings to fermions and gauge bosons were performed for CLIC for
some benchmark energies and luminosities.
Table~\ref{tab:higgs:params} shows the expected precision in the measurement of the Higgs couplings and was taken from~\cite{Sicking:2016zcl} (see~\cite{Sicking:2016zcl, Abramowicz:2013tzc}
for details). The $\kappa_{Hii}$ are defined as
\beq
\kappa_{Hii} =
  \sqrt{\frac{\Gamma_{Hii}^{BSM}}{\Gamma_{Hii}^{SM}}} \;,
\eeq
which at tree-level is just the ratio of the Higgs coupling in
the BSM model and the corresponding SM Higgs coupling.  We have called the three benchmarks scenarios $Sc1$ (350 GeV), $Sc2$ (1.4 TeV) and $Sc3$ (3.0 TeV).
In this table we can see the foreseen precisions
that are expected to be attained  for each $\kappa_{Hii}$. With these predictions we can now ask what is the effect on
the parameter space of each model presented in the previous
section. This in turn will tell us how much an extra component from either
a singlet (or more singlets) or a doublet contributes to the $h_{125}$
scalar boson. Clearly, if no new scalar is discovered
one can only set bounds on the amount of mixing resulting from the addition of extra fields. In the case of a CP-violating model it is possible
to set a bound on the ratio of pseudoscalar to scalar Yukawa couplings, where there is an important interplay with the results
from EDM  measurements. The results presented in this section always assume that the measured central value is the SM expectation,
meaning that all $\kappa_{Hii}$ in Table~\ref{tab:higgs:params} have a central value of 1. Small deviations from the central value will not
have a significant effect on our results because the errors are very small. If significant deviations from the SM predicted values are found
the data has to be reinterpreted for each model.

Starting with the simplest extension, the CxSM, there are either one or two singlet components that mix with the real neutral
part of the Higgs doublet. In the broken phase, where there are no dark matter
candidates, the admixture is given by the sum of the squared mixing matrix elements corresponding to the real and complex singlet parts, {\it i.e.}
\beq
\Sigma_i^{\text{CxSM}} = (R_{i2})^2 + (R_{i3})^2 \;,
\eeq
with the matrix $R$ defined in Eq.~(\ref{eq:rotsinglet}). If a dark matter candidate is present
one of the $R_{ij}, j=2,3$, is zero.  In any case the Higgs couplings to SM particles are all rescaled
by a common factor. Therefore, we just need to consider the most accurate Higgs coupling measurement
to get the best constraints on the Higgs admixture.
The maximum allowed singlet admixture  is given
by the lower bound on the best measured $\kappa$ value which at present is
\beq
\Sigma_{\text{max LHC}}^{\text{CxSM}} \approx 1 - \kappa_{\text{min}} \approx
11\% \;.
\eeq
In CLIC $Sc1$ the most accurate measurement is for the scaled coupling $\kappa_{HZZ}$, which
would give
\beq
\Sigma_{\text{max CLIC@350GeV}}^{\text{CxSM}}  \approx 0.85\% \;,
\eeq
while for $Sc3$ one would obtain, from  $\kappa_{HWW}$,
\beq
\Sigma_{\text{max CLIC@3TeV}}^{\text{CxSM}}  \approx 0.22\% \;.
\eeq
This implies, for this particular kind of extensions, that the chances of finding a new scalar are reduced due
to the orthogonality of the $R$ matrix. Note that in the limit of exact zero singlet component the singlet
fields do not interact with the SM particles. The results for a real singlet are similar, with the bound being exactly the same
but with a two by two orthogonal matrix replacing $R$. In this case it
is exactly the value 0.22\%  that multiplies
all production cross sections of the non-SM Higgs boson, after CLIC@3TeV.

\begin{figure}[h!]
  \centering
  \includegraphics[width=0.47\linewidth]{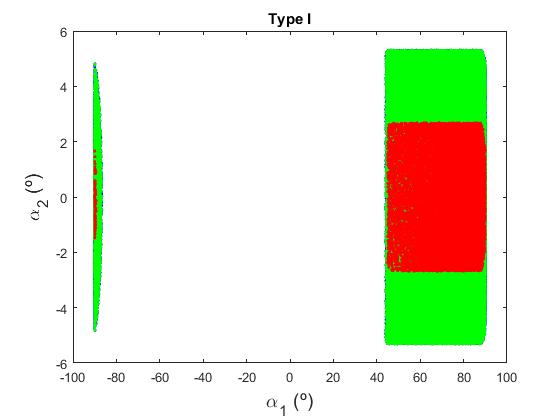}
  \includegraphics[width=0.47\linewidth]{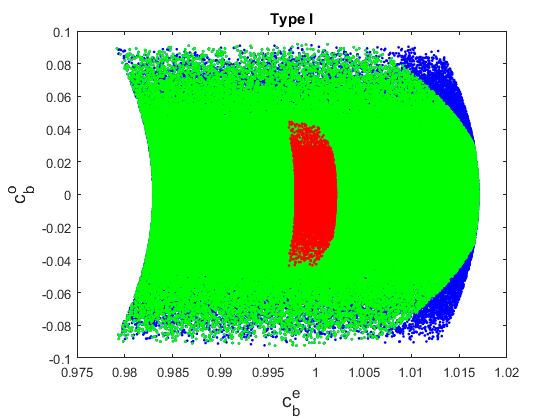}
  \caption{Mixing angles $\alpha_2$ vs. $\alpha_1$ (left) and $c_b^o$ vs. $c_b^e$ (right) for the C2HDM Type I.
    The blue points are for $Sc1$ but without the constraints from $\kappa_{Hgg}$ and  $\kappa_{H\gamma \gamma}$;
    the green points are for $Sc1$ including $\kappa_{Hgg}$ and the red points are for $Sc3$ including $\kappa_{Hgg}$ and  $\kappa_{H\gamma \gamma}$.
    }\label{fig:c2hdmI}
\end{figure}

We now discuss the C2HDM as this is the model with a CP-violating
scalar and one that shows a quite different behaviour in
the four independent Yukawa versions of the model. In fact, the constraints
act very differently in the four Yukawa versions of the model as shown in~\cite{Fontes:2017zfn}. This is particularly so
for the EDMs~\cite{Fontes:2017zfn} - while for Type II the electron EDM constraint almost
kills the pseudoscalar component of the the $bbH$ coupling, the same is not true for the Flipped model
and for the pseudoscalar component of the Higgs couplings to leptons in the Lepton Specific model.
Since different Yukawa couplings enter the two-loop Barr-Zee diagrams, a small EDM can either be the result of small CP-violating Yukawa couplings  or
come from cancellations between diagrams. This can even allow for maximally CP-violating Yukawa couplings of the $h_{125}$ in some cases~\cite{Fontes:2017zfn}.
%
%
So now the question is: in the long run, can CLIC give us relevant information that complements the one from EDMs?
How far can one expect to go in the knowledge of the Higgs nature by
putting together CLIC and EDM results, how well can
one constrain the CP-violating component of the 125 GeV Higgs boson?

In Fig.~\ref{fig:c2hdmI} (left) we present the mixing angles $\alpha_2$ versus $\alpha_1$ for the C2HDM Type I. The blue points are for $Sc1$ but without the constraints from
$\kappa_{Hgg}$ and $\kappa_{H\gamma \gamma}$; the green points are for $Sc1$ including $\kappa_{Hgg}$ (the measurement of $\kappa_{H\gamma \gamma}$ was not included because it is not available)
and the red points are for $Sc3$ including $\kappa_{Hgg}$ and
$\kappa_{H\gamma \gamma}$. Note that the  $\kappa_{Hgg}$ and
$\kappa_{H\gamma \gamma}$ are the only measurements of couplings that
can probe the interference
between Yukawa couplings (in the case of $\kappa_{Hgg}$) and between Yukawa and Higgs gauge couplings  (in the case of $\kappa_{H\gamma \gamma}$).
In the right panel
of Fig.~\ref{fig:c2hdmI} we show the pseudoscalar component of the $b$-quark Yukawa coupling $c_b^o$ versus its scalar component $c_b^e$. Because in Type I all Yukawa couplings
are equal, this plot is valid for all Type I Yukawa couplings. One can then expect, by the end of the CLIC operation, all pseudoscalar (scalar) Type I Yukawa couplings to be
less than roughly 5\% (0.5 \%) away from the SM expectation. We again stress that this result assumes that experiments will not see deviations from the SM.

Recently, in~\cite{Jeans:2018anq} a study was performed for a 250 GeV electron-positron collider for Higgsstrahlung
events in which the $Z$ boson decays into electrons, muons, or hadrons, and the Higgs boson decays into $\tau$ leptons, which
subsequently decay into pions. The authors found that for an integrated luminosity of 2 $ab^{-1}$, the mixing angle between the CP-odd
and CP-even components, defined as
\begin{equation}
{\cal L}_i = g \bar{\tau} \left[ \cos  \psi_{CP}  + i \gamma_5   \sin  \psi_{CP} \right]  \tau H_i \;, \label{eq:yu}
\end{equation}
could be measured to a precision of $4.3^o$ which means that this is the best bound if the central measured value of the angle is zero.
Their result is translated into our notation via
\beq
 \tan  \psi_{CP}^{\tau} =  \frac{c^o(H_i \bar \tau \tau)}{c^e(H_i \bar \tau \tau)}   \;. \label{eq:yu2}
\eeq
Taking into account the values in Fig.~\ref{fig:c2hdmI}  (right) we obtain bounds on $\psi_{CP}^{top} = \psi_{CP}^{bottom} = \psi_{CP}^{\tau}  $ , for Type I,
(by looking at the maxima and minima of each component in the plot)
that are of the order of
$6^o$ for CLIC@350GeV and $3^o$ for CLIC@3TeV. Therefore the indirect bounds are of the same order of magnitude as
the direct ones.

\begin{figure}[h!]
  \centering
  \includegraphics[width=0.47\linewidth]{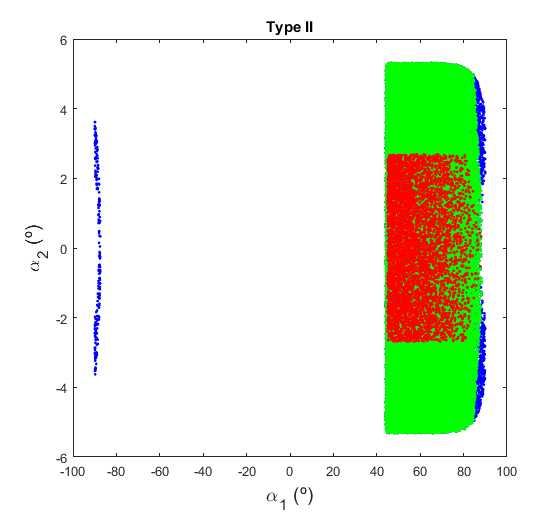}
  \includegraphics[width=0.47\linewidth]{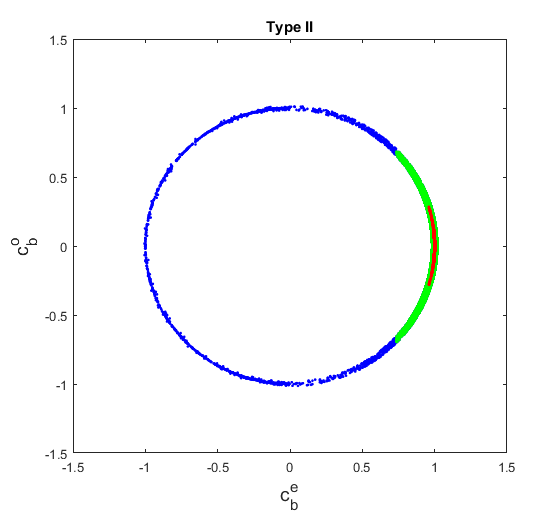}
  \caption{Mixing angles $\alpha_2$ vs. $\alpha_1$ (left) and $c_b^o$ vs. $c_b^e$ (right) for the C2HDM Type II.
    The blue points are for $Sc1$ but without the constraints from
    $\kappa_{Hgg}$ and $\kappa_{H\gamma \gamma}$;
    the green points are for $Sc1$ including $\kappa_{Hgg}$ and the red points are for $Sc3$ including $\kappa_{Hgg}$ and $\kappa_{H\gamma \gamma}$.
    }\label{fig:c2hdmII}
\end{figure}

In Fig.~\ref{fig:c2hdmII} (left) we present the mixing angles $\alpha_2$ vs. $\alpha_1$  for the C2HDM Type II. In the right panel
we again show the pseudoscalar component of the $b$-quark Yukawa coupling $c_b^o$ vs. its scalar component $c_b^e$.
The blue points are for $Sc1$ without the constraints from $\kappa_{Hgg}$ and  $\kappa_{H\gamma \gamma}$. These loop induced couplings
are the only ones where interference between Yukawa couplings and Higgs gauge couplings occur. Therefore, whatever the precision on the measurement of tree-level couplings is,
the result will always be a ring in that plane, that will become increasingly thiner with growing precision. However,
even for CLIC@350GeV, if the constraint for $\kappa_{Hgg}$ is included, the ring is reduced
to the green arch shown in the figure. By the end of the CLIC operation the arch will be further reduced to the red one. As discussed
in previous works, a very precise measurement of $\kappa_{Hgg}$ or  $\kappa_{H\gamma \gamma}$ will kill the wrong-sign limit\footnote{The wrong
sign limit refers to a Yukawa coupling that has a relative (to the
coupling of the Higgs boson to the massive gauge bosons) minus sign to the corresponding SM coupling~\cite{Ferreira:2014naa, Ferreira:2014dya}.},
which corresponds in the figure to $c_b^e = -1$.
Now, how do these bounds compare do the direct ones from $h_{125} \to  \tau^+ \tau^-$? In Type I the same bounds apply to all $\psi_{CP}$.
At the same time the bound on $\psi_{CP}^{top}$ is the same in all models and it was already discussed
for Type I.
In Type II $\psi_{CP}^{bottom} = \psi_{CP}^{\tau}$ and from Fig.~\ref{fig:c2hdmII} (right) we obtain bounds on $\psi_{CP}^{bottom} $ that are of the order of
$30^o$ for CLIC@350GeV and $15^o$ for CLIC@3TeV.
Therefore, we conclude that for Type II the indirect bounds cannot compete with the direct ones. The EDM constraints
also play a very important role in probing the CP-odd components of
the couplings. In fact, in the particular scenario of the Type II C2HDM
in which the lightest
Higgs boson is the 125 GeV
scalar, the bound is already constraining  $\psi_{CP}^{bottom} $ to be
below $20^o$~\cite{Fontes:2017zfn} clearly competing with the expectations for
CLIC.

The present best measurement for the electron EDM was obtained by the ACME collaboration, with an upper bound of
$|d_e|< 9.3 \times 10^{-29} \, e \, cm$ (90\% confidence)~\cite{Baron:2013eja}
and by the JILA collaboration with an upper bound of
$|d_e|< 1.3 \times 10^{-28} \, e \, cm$ (90\%
confidence)~\cite{Cairncross:2017fip}. ACME~II is expected to increase
the statistical sensitivity by an order of magnitude~\cite{Baron:2016obh}
 relative to the ACME~I result. There are several other planned experiments that could result in an increase in sensitivity by two to three orders of
 magnitude~\cite{1742-6596-691-1-012017, Yamanaka:2017mef}. These experiments together with the input from CLIC would certainly improve our knowledge on
 the nature of the Higgs boson.
%

%
%

\begin{figure}[h!]
  \centering
  \includegraphics[width=0.47\linewidth]{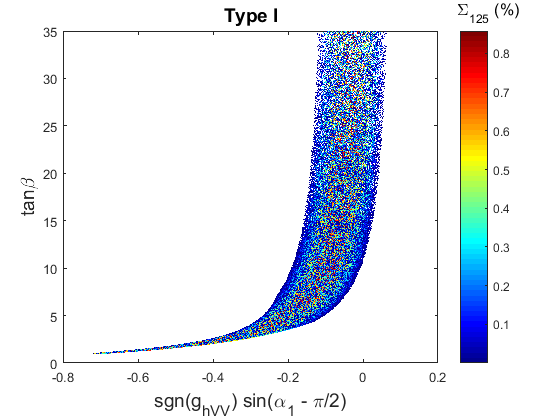}
  \includegraphics[width=0.47\linewidth]{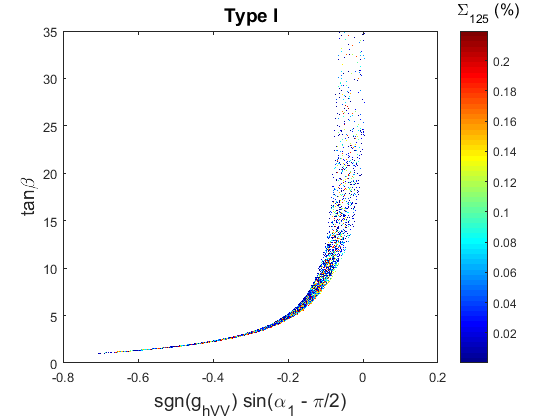}
  \caption{$\tan \beta$ as a function of $\sin (\alpha_1 - \frac{\pi}{2})$  for Type I in $Sc1$ (left) and $Sc3$ (right). The factor $- \frac{\pi}{2}$ is due to a different
  definition of the rotation angles relative to the 2HDM.  Also shown
  in the colour code is the amount of singlet admixture present in $h_{125}$.
    }\label{fig:n2hdmI}
\end{figure}
The predictions for the N2HDM are very similar to the ones for the 2HDM and we will discuss them together.
Although the N2HDM has an extra singlet field relative to the 2HDM, the couplings to gauge bosons and fermions are
very similar. For instance, for the lightest Higgs boson the couplings to massive gauge bosons are related via $g_{hVV}^{N2HDM} = \sin \alpha_2 \, g_{hVV}^{2HDM} $
which results in some extra freedom for the N2HDM parameter space. In Fig.~\ref{fig:n2hdmI} we show $\tan \beta$ as a function of $\sin (\alpha_1 - \frac{\pi}{2})$ for Type I in $Sc1$ (left) and $Sc3$ (right) (the lepton-specific case behaves very similarly).
The only notable difference
between the N2HDM and the 2HDM is the colour bar where we show the
percentage of the singlet component in the $125\,\text{GeV}$ Higgs boson,
$\Sigma_{125}= (R_{i3})^2$. In a previous work~\cite{Muhlleitner:2017dkd} we have shown that
before the LHC run 2 the allowed admixture of the singlet was below
25\% for Type I and the predictions for CLIC@350GeV and  CLIC@3TeV are
below 0.85\% and 0.22\%, respectively.

As expected, the allowed parameter space gets closer and closer to the
SM line, that is the line $\sin(\beta - \alpha) = 1$ (alignment limit). Note that unless one detects a new particle there is no way to find
the value of $\tan \beta$ if the models are  in the alignment
limit.  In fact, considering that the lightest Higgs boson is the 125 GeV one,
if we are in the alignment limit, $\sin(\beta -\alpha)=1$ in the 2HDM,\footnote{In the N2HDM, the alignment limit is attained for $\cos(\beta -\alpha_1) \cos \alpha_2=1$
(where the $\cos(\beta -\alpha_1)$ appears due to a different definition of the angle $\alpha_1$ relative to the 2HDM). This means the N2HDM has SM-like couplings
when $\cos(\beta -\alpha_1) = 1$ \textit{and} $ \cos \alpha_2=1$. In
this limit the 125 GeV Higgs boson has no contribution from the singlet field.}
all couplings of the 125 GeV Higgs boson to
the other SM particles are independent of the value of $\tan \beta$
(including the triple Higgs coupling). If the 125 GeV Higgs boson is not
the lightest scalar in the model, the limits change but the physics is the same.

\begin{figure}[h!]
  \centering
  \includegraphics[width=0.47\linewidth]{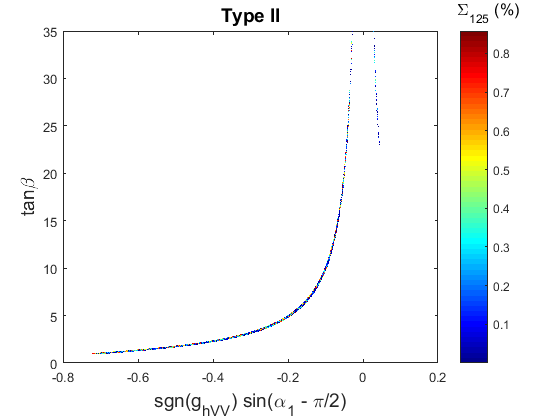}
  \includegraphics[width=0.47\linewidth]{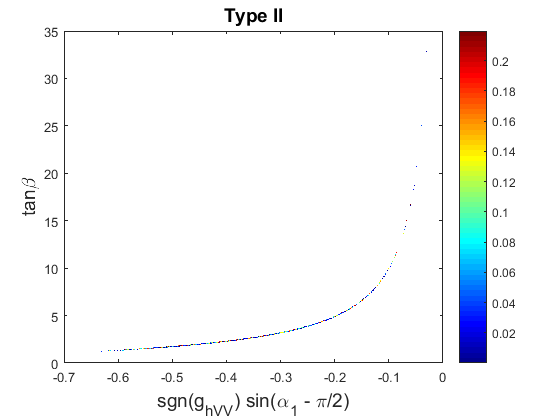}
  \caption{$\tan \beta$ as a function of $\sin (\alpha_1 - \frac{\pi}{2})$ for Type II in $Sc1$ (left) and $Sc3$ (right). The factor $- \frac{\pi}{2}$ is due to a different
  definition of the rotation angles relative to the 2HDM.  Also shown in the colour code is the amount of singlet present in $h_{125}$.
    }\label{fig:n2hdmII}
\end{figure}

In Fig.~\ref{fig:n2hdmII} we show $\tan \beta$ as a function of $\sin (\alpha_1 - \frac{\pi}{2})$ for Type II in $Sc1$ (left) and $Sc3$ (right). These are typical
plots not only for a Type II N2HDM but also for a Type II 2HDM (and very similar plots are obtained for the Flipped versions of both models). As previously
discussed we see that the right leg, corresponding to the wrong-sign limit, is very dim in the left plot and vanishes in the right plot. Again, this is true for
both the 2HDM and the N2HDM. As for the percentage of the singlet
component, it was constrained to 55\% for Type II N2HDM at the end of run 1~\cite{Muhlleitner:2017dkd}
and the predictions for CLIC@350GeV and  CLIC@3TeV are below about 0.8\% and 0.2\%, respectively.

%
\begin{figure}[h!]
  \centering
  \includegraphics[width=0.47\linewidth]{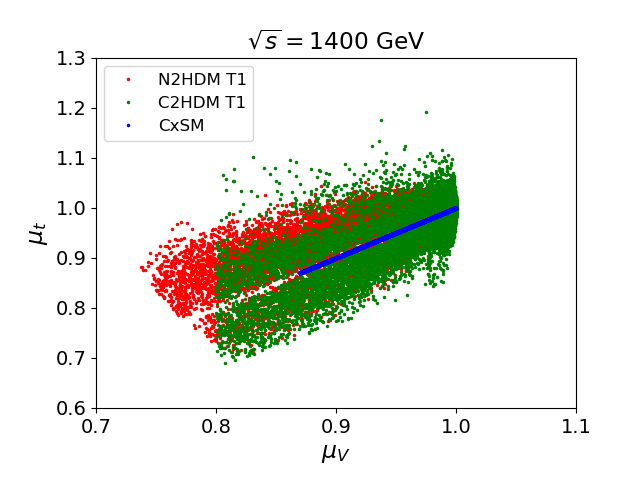}
  \includegraphics[width=0.47\linewidth]{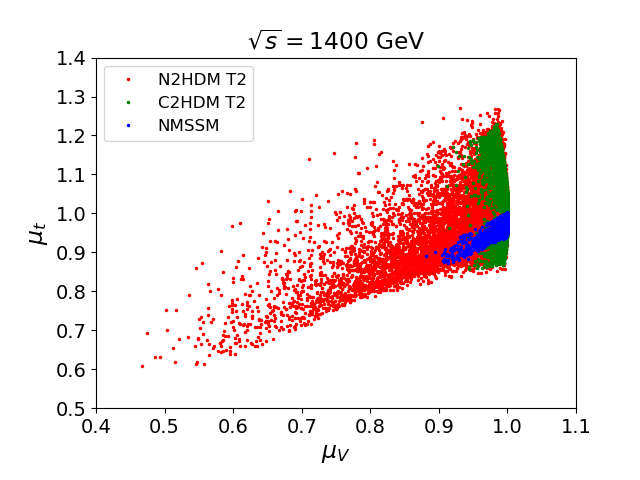}
  \caption{$\mu_{t} = \sigma_{\bar tth}^{BSM}/\sigma_{\bar tth}^{SM}$ as a function of $\mu_{V} = \sigma_{VVh}^{BSM}/\sigma_{VVh}^{SM} =  \left( g_{VVh}^{BSM}/g_{VVh}^{SM} \right)^2$,
  where $V=W, Z$.
   for the 2HDM and N2HDM Type I and the CxSM (left) and for the 2HDM
   and N2HDM Type II and the NMSSM (right) for 1.4 TeV.
    }\label{fig:mutth2}
\end{figure}
%

We end this section with a discussion on the correlations between different cross section measurements for the different
models. In Fig.~\ref{fig:mutth2} we present $\mu_{t} =
\sigma_{\bar tth}^{BSM}/\sigma_{\bar tth}^{SM}$ as a function of $\mu_{V} =
\sigma_{VVh}^{BSM}/\sigma_{VVh}^{SM} = \left( g_{VVh}^{BSM}/g_{VVh}^{SM} \right)^2$
 for the 2HDM and N2HDM Type I and the CxSM (left) and for the 2HDM
 and N2HDM Type II and the NMSSM (right) for 1.4 TeV, including the present LHC coupling constraints. We can find in the plots distinct regions where precise measurements
 that deviate from the SM prediction could hint on a specific model. Take for instance the plot on the right and let us assume that the $\mu$'s could be measured with 5\% precision. In this
 case a measurement $(\mu_{t},\, \mu_{V} ) = (1,\, 0.85)$ indicates that the model cannot be the C2HDM Type II nor the NMSSM. A measurement $(\mu_{t},\, \mu_{V} ) = (1.2,\, 1.0)$ excludes
 the NMSSM but not the remaining two models, in their Type II versions.
%
\begin{figure}[h!]
  \centering
\includegraphics[width=0.47\linewidth]{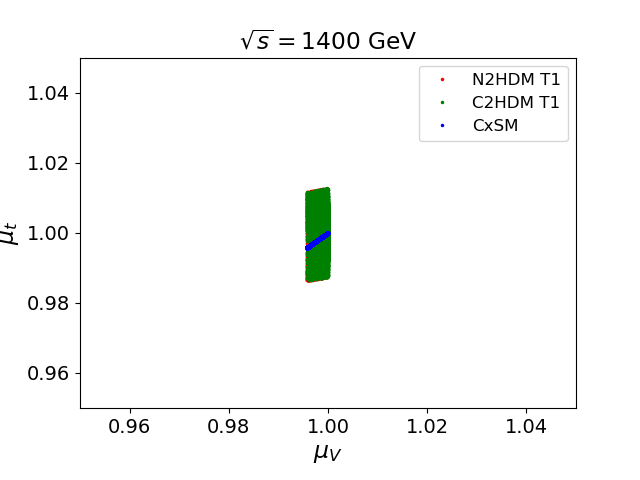}
\includegraphics[width=0.47\linewidth]{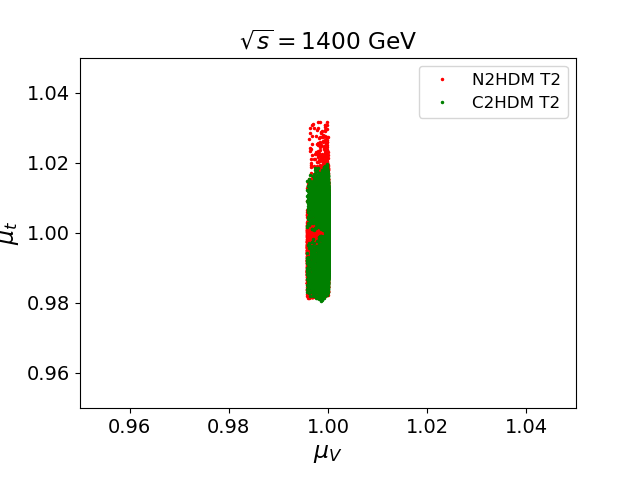}
  \caption{Same as Fig.~\ref{fig:mutth2}, but after imposing the constraints on the Higgs couplings from CLIC@350GeV.
    }\label{fig:mutth1}
\end{figure}
%


Finally, Fig.~\ref{fig:mutth1} is the same as Fig.~\ref{fig:mutth2} with the extra constraint of imposing
the bounds coming from the CLIC@350GeV run. The results from the 350 GeV run turn out to be so restrictive
that the allowed parameter space is heavily reduced in all models. In particular, all points of the NMSSM are
excluded, considering that the measurements have the SM central values and no new physics was found~\footnote{Note that the SM-like limit is only attained for vanishing singlet admixtures. }.
The behaviour is very similar for all models and  in this case a deviation from the SM expectation could exclude some
models.  However, since we are already at the \% level electroweak radiative corrections would have to be taken
into account for the different models.
Note that because $e^+ e^- \to \bar t t h$ (for which both Yukawa couplings and Higgs gauge couplings contribute)
is not kinematically allowed for 350 GeV, the study of the correlations between this process
and associated or $W$-fusion cross sections (for which only Higgs
gauge couplings contribute) can only be performed for 1.4 TeV.

\section{Signal Rates of the non-SM-like Higgs Bosons \label{sec:sigrates}}
In this section we present and compare the rates of the neutral non-SM-like Higgs bosons
in the most relevant channels at a linear collider. We denote by  $H_\downarrow$ the lighter and by
$H_\uparrow$ the heavier of the two neutral non-$h_{125}$ Higgs bosons.
All signal rates are obtained by multiplying the production cross section with the corresponding
branching ratio obtained from {\tt sHDECAY}, {\tt C2HDM\_HDECAY}, {\tt
  N2HDECAY} and {\tt NMSSMCALC}.
For the particular processes presented in this section, there is no distinction between particles
with definite CP-numbers and CP-violating ones and they are therefore treated on equal footing.
The main production processes for a Higgs boson at CLIC are associated production with
a $Z$ boson, $e^+ e^- \to Z H_i$, and $W$-boson
fusion $e^+ e^- \to \nu \bar \nu H_i$. We will be presenting
results for two centre-of-mass energies, $\sqrt{s}=350$ GeV and $\sqrt{s}=1.4$ TeV. In the case of the former
the cross sections are comparable in the mass range presented while for the latter the $W$-boson
fusion cross section dominates in the entire Higgs boson mass range.
In order to give some meaning to the event
rates presented in this section, we will use as a rough reference that at CLIC 10$^{-1}\,$fb for $Sc1$ correspond to 50 signal events and 10$^{-2}\,$fb for $Sc2$ correspond to 150 signal events.

\subsection{The 350 GeV CLIC}
\begin{figure}[h!]
  \centering
  \includegraphics[width=0.47\linewidth]{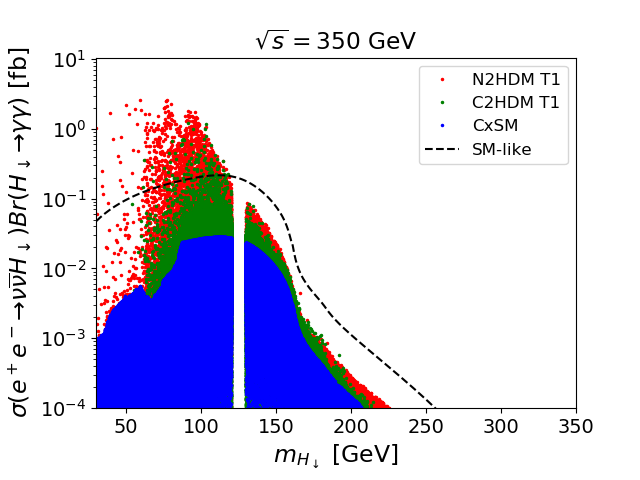}
  \includegraphics[width=0.47\linewidth]{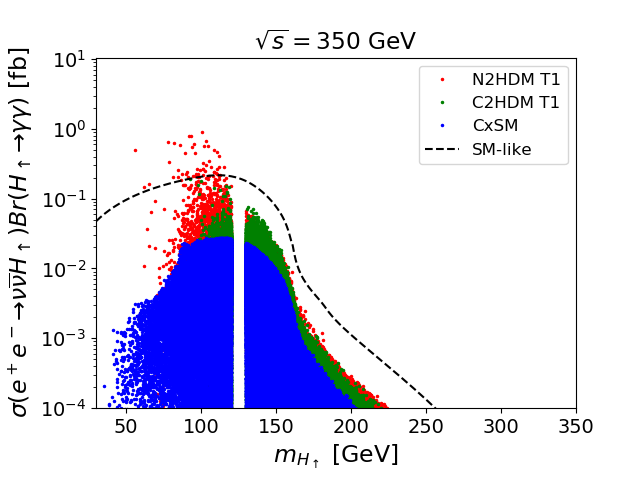}
  \caption{Total rate for $e^+ e^- \to \nu \bar \nu H_{i} \to  \nu
    \bar \nu \gamma \gamma$ as a function of the Higgs boson mass for $\sqrt{s}=350$ GeV. The models presented
    are the CxSM and the Type I versions of the N2HDM and C2HDM.
  Also shown is the line for a SM-like Higgs boson. On the left panel
  we present the results for the lighter Higgs boson,
  $H_{\downarrow}$, and on the right we show the results for the
  heavier Higgs boson, $H_{\uparrow}$.
    }\label{fig:t1380}
\end{figure}

In Fig.~\ref{fig:t1380} we present the total rate for $e^+ e^- \to \nu
\bar \nu H_{i} \to  \nu \bar \nu \gamma \gamma$ as a function of the
Higgs boson mass for the CxSM and for the Type I versions of the
N2HDM and C2HDM.
  Also shown is the line for a SM-like Higgs boson. On the left panel
  we present the results for the lighter Higgs boson,
  $H_{\downarrow}$, and on the right we show the results for the
  heavier Higgs boson,
  $H_{\uparrow}$. The trend shown in the two plots is the same for all other final states. There is a hierarchy with the points of the N2HDM reaching the largest cross sections followed closely by the C2HDM
  and finally by the CxSM. This is easy to understand since the CxSM
  is the model with the least freedom - all couplings of the Higgs boson to SM particles are modified by the same factor - while the N2HDM is the least constrained
  model. This means that it is possible to distinguish between the singlet and the Type I doublet versions if a new scalar is found with a large enough rate.
  The $\gamma \gamma$ final state is one where the branching ratio decreases very fast with the mass. Still it is clear that there are regions of the parameter space that have large enough
  production rates to be detected at the 350 GeV CLIC.  We would like to stress that the behaviour seen in the plots regarding the event rates for the lighter (left) and for the heavier (right) scalar
  is the same for the remaining final states and we will only show
  plots for the lighter Higgs boson in the remainder of this section.

\begin{figure}[h!]
  \centering
  \includegraphics[width=0.47\linewidth]{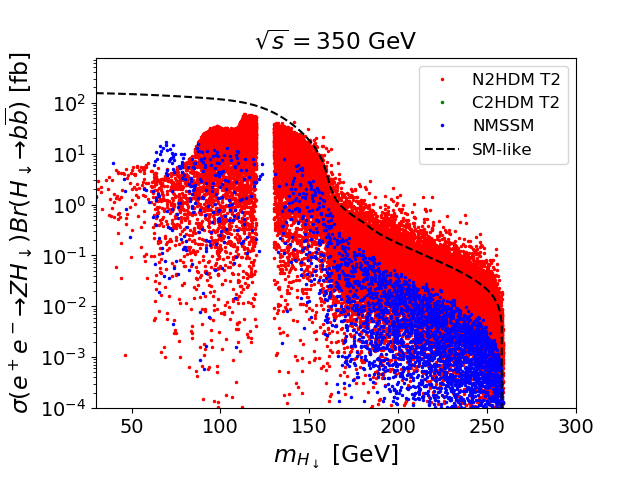}
  \includegraphics[width=0.47\linewidth]{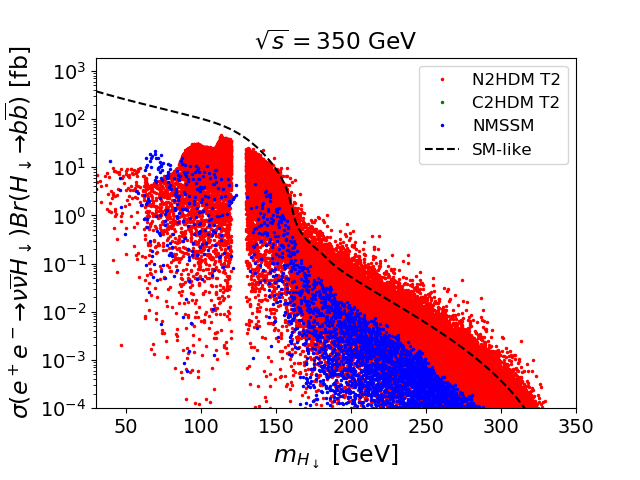}
  \caption{ Total rate for $e^+ e^- \to Z H_{\downarrow} \to  Z b \bar b$ (left) and for
  $e^+ e^- \to \nu \bar \nu H_{\downarrow} \to  \nu \bar \nu b \bar b $ (right) as a function of  $m_{H_{\downarrow}}$ for $\sqrt{s}=350$ GeV.
  Plots are shown for the NMSSM and for the Type II versions
     of the N2HDM and C2HDM.   Also shown is the line for a SM-like Higgs boson.
    }\label{fig:t2380}
\end{figure}

In Fig.~\ref{fig:t2380} we present the total rate for $e^+ e^- \to Z H_{\downarrow} \to  Z b \bar b$ (left) and for
  $e^+ e^- \to \nu \bar \nu H_{\downarrow} \to  \nu \bar \nu b \bar b $ (right) as a function of  $m_{H_{\downarrow}}$
  for $\sqrt{s} = 350$ GeV, for the NMSSM and for the Type II versions
     of the N2HDM and C2HDM. Clearly there is plenty of parameter space to be explored in the NMSSM and
     even more in the Type II N2HDM. For the Type II C2HDM, as discussed in a previous work~\cite{Fontes:2017zfn},
     the constraints are such that points with masses below about 500 GeV are excluded. Again there are regions
     where the models can be distinguished but not if the cross sections are too small. As expected, for this
     centre-of-mass energy there is not much difference between the two production processes (for instance for a 125 GeV scalar
     $\sigma (e^+ e^- \to Z H_i) = \sigma (e^+ e^- \to  \nu \bar \nu H_i) $ for $\sqrt{s} \approx 400$ GeV; as the scalar mass grows
     so does the energy for which the values of the cross sections cross). We have also
     checked that the behaviour of the total rates does not change
     significantly when the Higgs boson decays to other
     SM particles. That is, although the rates are much higher in $H_i
     \to b \bar b$ than in $H_i \to \gamma \gamma$, the overall behaviour is the same.
The highest rates are
     obtained in all models for the final states $b \bar b$, $W^+W^-$, $ZZ$
     and $\tau^+ \tau^-$.

\subsection{The 1.4 TeV CLIC}
\begin{figure}[h!]
  \centering
  \includegraphics[width=0.47\linewidth]{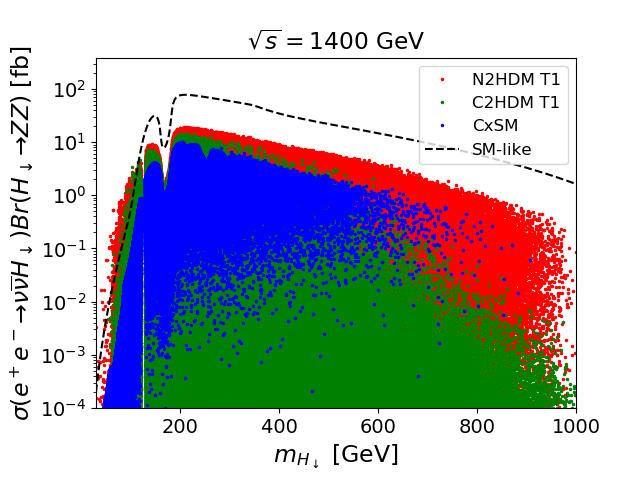}
  \includegraphics[width=0.47\linewidth]{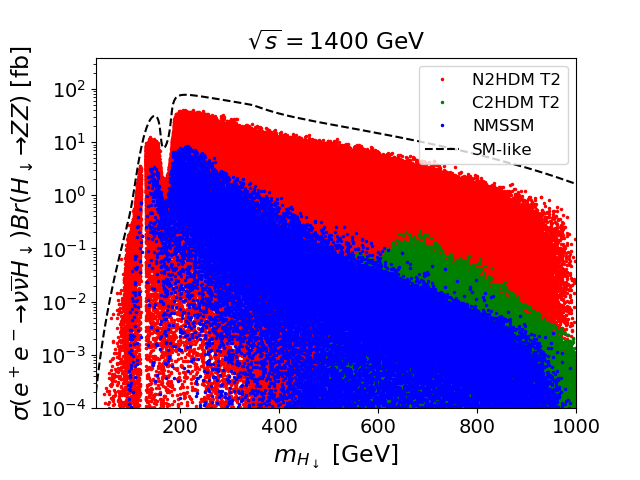}
  \caption{Total rate for  $e^+ e^- \to \nu \bar \nu H_{\downarrow}
    \to  \nu \bar \nu ZZ $ as a function of the lighter Higgs
    boson mass for $\sqrt{s}=1.4$ TeV.
  Left: models CxSM and Type I N2HDM and C2HDM; right: NMSSM and Type
  II N2HDM and C2HDM. Also shown is the line for a SM-like Higgs
  boson.
    }\label{fig:1500p1}
\end{figure}
As the centre-of-mass energy rises the $W$-fusion process becomes the dominant one. In Fig.~\ref{fig:1500p1}
we present the total rate for  $e^+ e^- \to \nu \bar \nu H_{\downarrow} \to  \nu \bar \nu ZZ $ as a function of the lighter Higgs mass for $\sqrt{s}=1.4$ TeV.
In the left panel we show the rates for the CxSM and for the Type I N2HDM and C2HDM while in the right panel plots for
the NMSSM and the Type II N2HDM and C2HDM are shown. We can expect that total rates above roughly $10^{-2}\,$fb can definitely be explored at CLIC@1.4TeV.
Hence, all models can be explored in a very large portion of the parameter space and again there are regions where the models
are clearly distinguishable. The plots do not present any major differences when we change the final states as previously discussed.
\s

\begin{figure}[h!]
  \centering
  \includegraphics[width=0.47\linewidth]{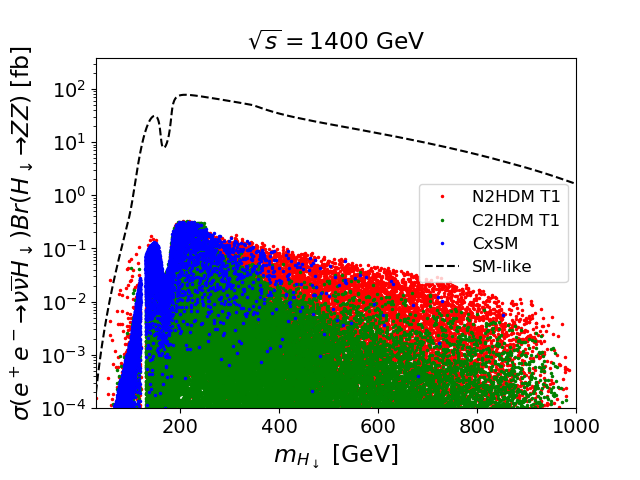}
  \includegraphics[width=0.47\linewidth]{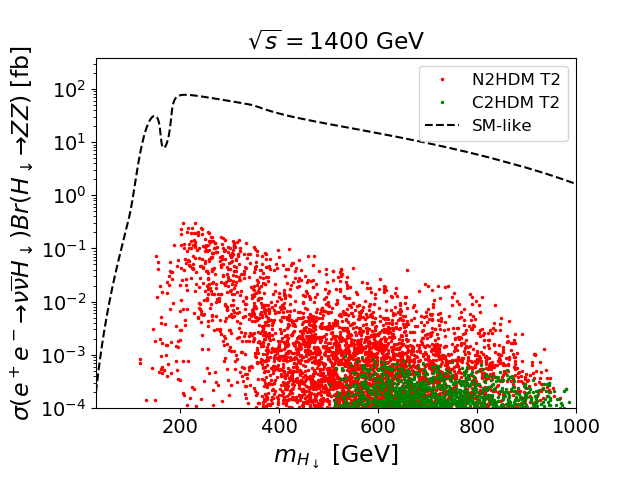}
  \caption{Same as figure~\ref{fig:1500p1}  after imposing the final results for the 350 GeV run.
    }\label{fig:1500p2}
\end{figure}
%
However, once the 350 GeV run is complete, even if no new scalar is found, the measurement of the 125 GeV Higgs couplings
will be increasingly precise which in turn reduces the parameter space of the model. In Fig.~\ref{fig:1500p2}
we present the total rate for  $e^+ e^- \to \nu \bar \nu
H_{\downarrow} \to  \nu \bar \nu ZZ $ as a function of the lighter
Higgs boson mass for $\sqrt{s}=1.4$ TeV
(same as Fig.~\ref{fig:1500p1}) but where we have included the predictions on the Higgs coupling measurements
after the end of the 350 GeV run.  We see that after imposing the constraints on the Higgs couplings the cross sections decrease by
more than one order of magnitude. We find that the models can all be probed but are no longer distinguishable
just by looking at the total rates to SM particles.
 Interestingly, all points from the NMMSM disappear when we impose
the constraints from the 350~GeV run. This is of course related to the fact that we have used the SM central values
for all predictions but it could very well be that at the end of this run we could be celebrating the discovery
of a new NMSSM particle --- or from any other model!

\begin{figure}[h!]
	\centering
	\includegraphics[width=0.47\linewidth]{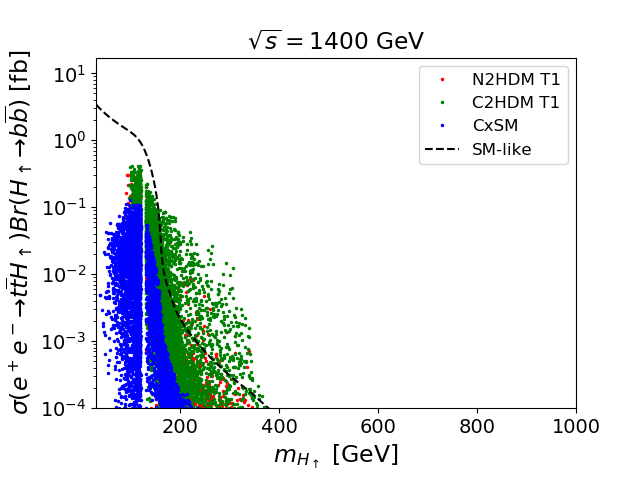}
	\includegraphics[width=0.47\linewidth]{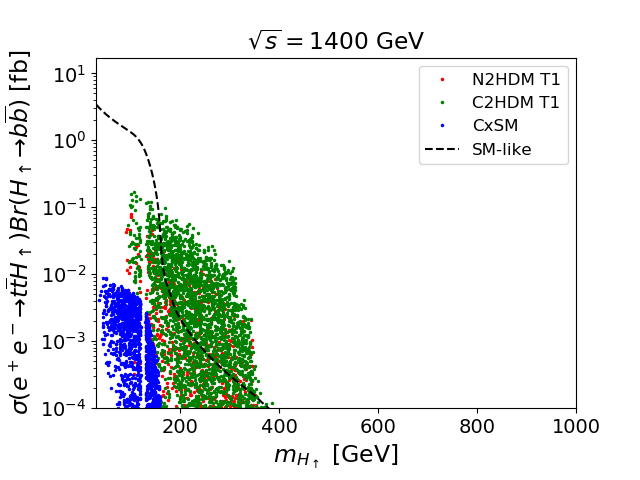}
	\caption{Total rates for $e^+e^-\to t\bar{t}H_\uparrow \to t\bar{t}b\bar{b}$ for the type 1 N2HDM and C2HDM and CxSM. No 350 GeV CLIC constraints (left) and with constraints (right).}\label{fig:1500p3}
\end{figure}

In Fig.~\ref{fig:1500p3} we also include this comparison for $t\bar{t}H$
production with (right) and without (left) the 350 GeV run constraints. Apart
from the CxSM --- where there is a common scaling of all Higgs couplings --- the
constraints from the 350~GeV run have a much smaller impact on the $t\bar{t}H$
cross section than on the gauge-boson mediated processes. This happens because a
$h_{125}$ Yukawa coupling close to one does not require the Yukawa couplings of
the other Higgs bosons to be small. The resulting $t\bar{t}H$ cross sections in
the N2HDM and C2HDM can indeed be comparable or even larger than the $\nu\bar{\nu}H$
cross section. Therefore, $t\bar{t}H$ production becomes a highly relevant
search channel if no additional Higgs bosons are discovered during the 350~GeV
run.

\section{Conclusions \label{sec:concl}}

We have investigated extensions of the SM scalar sector in several specific models:  the CxSM, the 2HDM, C2HDM and N2HDM in the Type I and Type II versions as well as the NMSSM.
The analysis is based on three CLIC benchmarks with centre-of-mass energies of 350 GeV, 1.4 TeV and 3 TeV.
For each benchmark run, the precision in the measurement of the Higgs couplings was used to study possible deviations from the -- CP-even and doublet-like -- expected behaviour of the discovered Higgs boson.
We concluded that the constraints on the admixtures
of both a singlet and a pseudoscalar component to the $125\,\text{GeV}$ Higgs boson, improve substantially from tens of percent to well below 1\%  when going from the LHC to the last stage of CLIC.
In fact, as shown in~\cite{Muhlleitner:2017dkd}, after the LHC Run 1 the constraints on the admixtures were as shown in table~\ref{tab:admixtures},
where $\Sigma$ stands for the singlet admixture and $\Psi$ is the pseudoscalar admixture.
As noted in~\cite{Muhlleitner:2017dkd} the upper bound on $\Psi$ for the C2HDM type II is mainly due to the EDM constraints. \s
\begin{table}[h!]
\begin{center}
\begin{tabular}{lcccccc} \toprule
Model & CxSM & C2HDM II & C2HDM I & N2HDM II & N2HDM I & NMSSM \\ \midrule
$\left(\Sigma\, {\rm or} \,\Psi\right)_{\text{allowed}}$ & 11\% & 10\% & 20\% & 55\% & 25\% &
 41\% \\
\bottomrule
\end{tabular}
\caption{Allowed singlet and pseudoscalar (for the C2HDM)
  admixtures. \label{tab:admixtures}}
\end{center}
\end{table}

With the CLIC results the limits on the admixtures are completely dominated by the measurement of $ \kappa_{HZZ}$
for $Sc1$ and by $\kappa_{HWW}$ for $Sc2$ and $Sc3$ through the unitarity relation
\begin{equation}
  \kappa^2_{ZZ,WW} + \Psi / \Sigma \leq 1\,
\end{equation}
where the sum rule includes the factor $R_{i3}$, which is either the pseudoscalar, or the singlet component depending on the model.
Since this holds in all our models the constraints become independent
of both model and Yukawa type and are given by
\begin{itemize}
  \item $Sc1$:   $\Sigma, \Psi < 0.85\%$ from $\kappa_{HZZ}$
  \item $Sc2$: $ \Sigma, \Psi < 0.30\%$ from $\kappa_{HWW}$
  \item $Sc3$: $ \Sigma, \Psi < 0.22\%$ from $\kappa_{HWW}$
\end{itemize}

In the second part of this work we investigated the potential to discover and study additional Higgs bosons at CLIC in $W$-boson fusion and Higgsstrahlung.
We checked whether the models could be distinguished by a discovery in the first stage of CLIC.
If no New Physics is found in the first stage of CLIC we discussed if the parameter space of the models still allows for large enough
rates to be probed at the second stage.
\begin{itemize}

\item
As expected the results are very similar for $W$-fusion and Higgsstrahlung for $\sqrt{s} = 350$ GeV. For the other two benchmark energies
the $W$-fusion process dominates. Since the difference relative to the SM in both production processes is in the coupling $hVV$, $V=W,\, Z$,
even for $\sqrt{s} = 350$ GeV, where the cross sections are of the same order, the two processes give the same information about the models.

\item
For  $\sqrt{s} = 350$ GeV and for Type I models and CxSM, the latter is always the most constrained model
as the couplings of the Higgs boson to SM particles are all modified by the same factor. Hence the Type I N2HDM
and C2HDM, which in most cases are barely distinguishable, have rates that are always larger than the CxSM ones.
For some final states the N2HDM rates are slightly above the C2HDM ones but always below the SM-like line,
except for the $\gamma \gamma$ final states and only for Higgs boson masses below about 120 GeV. In these
Type I models there are charged Higgs contributions in the $H_i \to \gamma \gamma$ loops and the charged
Higgs mass is not as constrained as in the Type II models.

\item
For  $\sqrt{s} = 350$ GeV and for Type II models and NMSSM, the C2HDM does not take part in the analysis
due to the constraint on the non-$125\,\text{GeV}$ Higgs boson as
previously explained. The Type II N2HDM has rates that are
always above the corresponding NMSSM ones. So, it is possible to distinguish the two models in several regions
of the parameter space which is expected since the N2HDM has more freedom.

\item
For  $\sqrt{s} = 350$ GeV and for Type II models and NMSSM, the heavier neutral scalar
can only be probed in the N2HDM where the rates can be up to two
orders of magnitude above the SM line (these plots were not shown).
CLIC can probe the lighter neutral scalar
boson in both the NMSSM and the N2HDM and distinguishing the two models based on total rates alone may be possible.

\item
For  $\sqrt{s} = 1400$ GeV the results are very similar in what regards the
relative rates for the different processes. The main difference comes from
imposing the predicted results for the 350 GeV run, if nothing is found and
using the SM prediction as central value. This constrains the admixtures --- and
by unitarity the gauge couplings of the non-SM-like Higgs bosons --- to tiny
values identical in all models. Therefore, the models become harder to
distinguish. Furthermore, due to the reduced gauge couplings $t\bar{t}H$ becomes
an important search channel for non-SM-like Higgs bosons.

\end{itemize}

Finally one should mention that as all predictions for the different models reach and go below the \% level, electroweak radiative
corrections come into play. As decoupling is present in all models there are certainly regions of the parameter space where
the tree-level results are close to the one-loop corrected ones. Still, we should make clear that already for CLIC@350GeV
we will reach a level of precision where no result is truly meaningful without the inclusion of electroweak radiative corrections.

\subsubsection*{Acknowledgments}
Special thanks to Philipp Basler for providing us with up-to-date NMSSM samples. We thank Roberto Franceschini for discussions.
We acknowledge the contribution of the research training group GRK1694 'Elementary particle physics at highest energy and highest precision',
for our meetings in Lisbon and in Karlsruhe.
PF and RS are supported in part by the National Science Centre, Poland, the
HARMONIA project under contract UMO-2015/18/M/ST2/00518.
JW gratefully acknowledges funding from the PIER Helmholtz Graduate School.
\vspace*{0.5cm}

\vspace*{1cm}
\bibliographystyle{h-physrev}
\bibliography{epluseminusarxivv3}

\end{document}